\documentclass{article}
\usepackage{arxiv}

\usepackage{graphicx} 
\usepackage{amsmath}
\usepackage{booktabs} 
\usepackage{adjustbox}
\usepackage[authoryear,round]{natbib}
\setcitestyle{authoryear,round,semicolon}
\usepackage{multirow}
\usepackage{makecell}
\usepackage{placeins} 
\usepackage{hyperref}
\renewcommand{\thefootnote}{\fnsymbol{footnote}}

\title{Representing the Surface Ocean in ECMWF’s data-driven forecasting system AIFS}
\author{
\begin{tabular}[t]{p{0.35\textwidth}p{0.35\textwidth}}
\centering Sara Hahner\thanks{Equal contribution} &
\centering Lorenzo Zampieri\footnotemark[1]
\end{tabular} \and
Jean-Raymond Bidlot \and
Philip Browne \and 
Matthew Chantry \and 
Mariana C. A. Clare \and 
Harrison Cook \and 
Peter Dueben \and 
Rachel Furner \and 
Sarah Keeley \and 
Josh Kousal \and 
Simon Lang \and 
Christian Lessig \and 
Gert Mertes \and
Kristian Mogensen \and 
Gabriel Moldovan \and
Charles Pelletier \and
Florian Pinault \and 
Ana Prieto Nemesio \and
Baudouin Raoult \and 
Irina Sandu \and 
Mario Santa Cruz \and 
Jakob Schloer \and 
Steffen Tietsche \and 
Hao Zuo}

\begin{document}

\maketitle

\setcounter{footnote}{0}
\renewcommand{\thefootnote}{\arabic{footnote}}

Machine-learning (ML) models, such as the  Artificial Intelligence Forecasting System (AIFS) at the European Centre for Medium-Range Weather Forecasts (ECMWF), have revolutionised weather forecasting in recent years. 
We present an extension of the AIFS that jointly models the atmosphere and surface ocean, including ocean waves and sea ice. 
The primary objective of this extension is to enhance machine-learning medium-range forecasting and enable new use cases by expanding the weather state to better capture coupled surface processes.
Our approach departs from traditional numerical models by not having two separate models for the atmosphere and marine components.
The joint model instead learns correlations across the entire atmosphere-ocean interface in a component-agnostic way. 
That way it can exploit the expressive capacity of modern ML architectures to learn cross-component relationships directly from the data.
Additionally, this approach avoids assumptions on the processes and the technical complexities of the coupling itself.

For training the model, we leverage tailored and targeted datasets, including ERA5, the ORAS6 ocean reanalysis, and an atmosphere-forced wave hindcast. 
We solve model design challenges such as missing values over land, multi-scale temporal dynamics, and physical realism of forecast fields and demonstrate the utility of loss scaling in guiding the learning process.
We evaluate how representing the surface ocean affects medium-range weather forecasts. 
We also assess the model’s ability to predict surface-ocean fields, including wave swell and tropical-cyclone cold wakes.
For nearly all evaluated marine variables, we observe an improvement of approximately one day in forecast skill at medium-range lead times compared to physics-based models. 
Furthermore, we demonstrate that the model is robust to idealised initial conditions outside the training distribution and responds to them in a physically consistent way.
Overall, our findings suggest that the joint AIFS modelling approach offers significant potential for combined atmosphere–ocean forecasting.
Our work provides a solid foundation for future development of data-driven coupled Earth system models with greater flexibility and physical fidelity.

\section{Introduction}

The surface ocean, ocean waves, and sea ice are key components of the Earth system, directly interacting with the atmospheric boundary layer. 
They govern air–sea exchanges of energy and freshwater while underpinning marine operations. 
Accurately forecasting the evolution of the surface ocean is therefore essential both for safe and efficient decision-making in marine environments and for achieving physically consistent, skilful weather prediction within the coupled Earth system.

In recent years, machine-learning (ML) models have shown strong performance in atmospheric forecasting, particularly at medium-range timescales, where they routinely outperform numerical models in both deterministic and probabilistic headline scores.
At the same time, they operate at a fraction of the computational resources. 
These ML models typically treat the atmosphere in isolation, without explicitly representing other Earth system components such as the ocean, land, or sea ice \citep{Pathak2022,Keisler2022,Lam2023,Chen2023,Bi2023,Lang2024a}. 
Nevertheless, the models still benefit indirectly from the influence of unresolved components.
This information is embedded, to a certain extent, in the training datasets (e.g. the Copernicus Climate Change Service ERA5 reanalysis \citep{Hersbach2020}) through both explicit ocean forcing in the forecasting system used to produce the reanalysis and implicit information provided by assimilated atmospheric observations. 

In traditional numerical prediction systems, representing marine–atmosphere interactions through coupled processes is well known to improve forecast skill \citep{Graham2005, Beraki2015}. 
While the first coupled ML models are being developed (e.g., \citet{Clark2024, CresswellClay2025, Duncan2025}), there has been very limited research on ML approaches that jointly model the atmosphere and the marine components---here defined as the surface ocean (including ocean waves) and sea ice.

This work aims to address this gap for two reasons. 
First, the gap hinders the development of fully integrated, data-driven Earth system models that are needed to better support applications such as coastal risk management, marine operations, and climate adaptation. 
Second, it remains unclear whether the implicit representation of marine processes in atmosphere-only ML models is sufficient to sustain forecast skill across long  
lead times, spatial scales, and oceanic regimes. 
Determining where, and on which temporal and spatial scales, an explicit marine representation becomes necessary remains an open research question.
This study investigates the impact of this explicit representation at medium-range timescales.

\begin{figure}
    \centering
    \adjustbox{center}{%
        \includegraphics[width=1.0\linewidth]{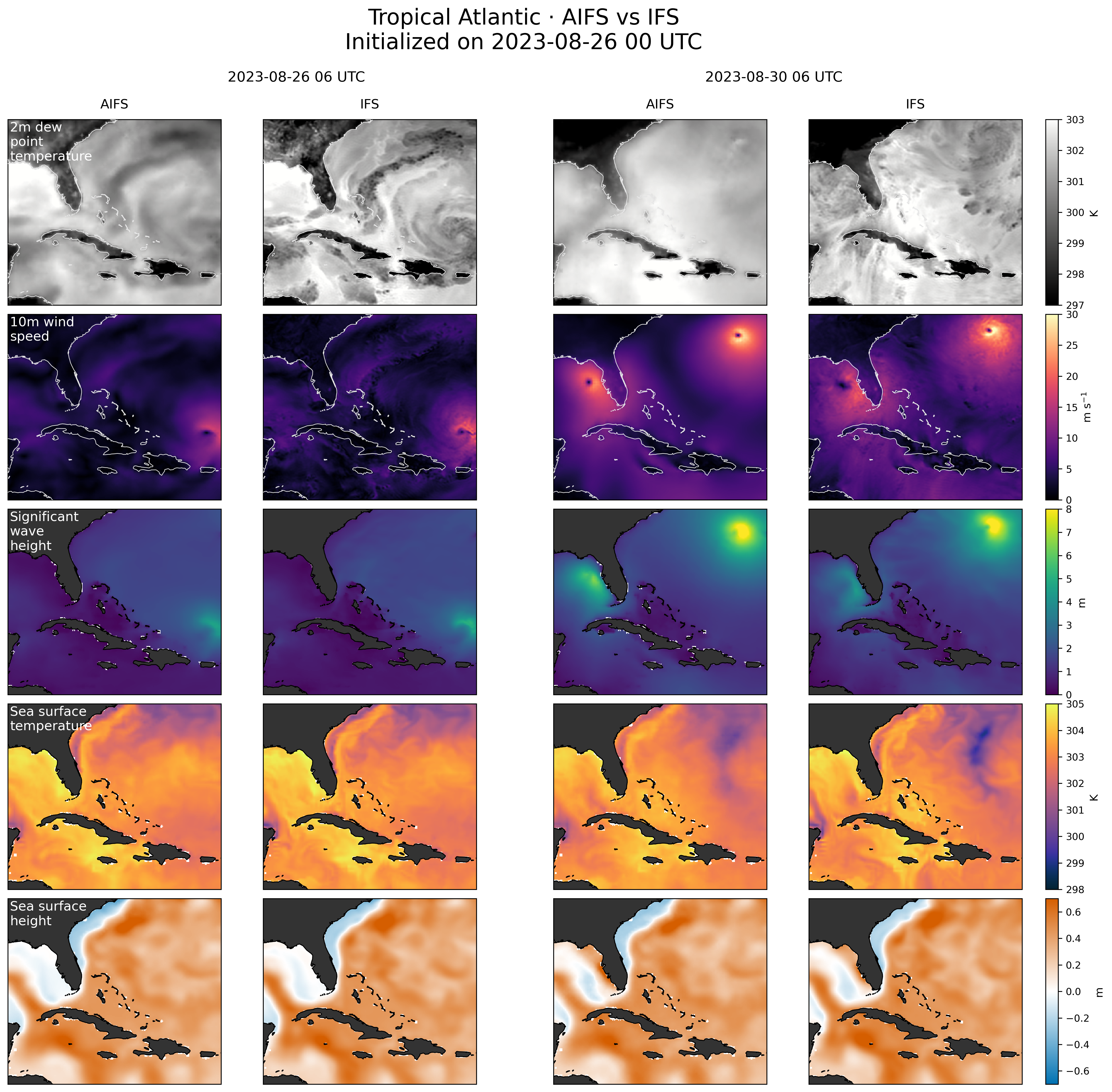}
    }
    \caption{
    Comparison of AIFS Marine and IFS forecasts of Hurricanes Idalia and Franklin over the Gulf of Mexico and the western North Atlantic, from forecasts initialised on 26 August 2023 at 00 UTC. 
    Columns show forecasts valid at +12\,h and +36\,h, with AIFS and IFS displayed side by side for each lead time. 
    Rows show (from top to bottom) 2\,m dew point temperature, 10\,m wind speed, significant wave height, sea surface temperature, and sea surface height.
    The figure highlights the alignment between atmospheric and marine responses, with strong near-surface winds co-located with enhanced wave activity and consistent ocean surface signals. 
    Spatial patterns are broadly comparable between AIFS Marine and IFS, illustrating the ability of the joint model to represent atmosphere-ocean interaction.
    }
    \label{fig:tc_AIFS_IFS}
\end{figure}

Previous work has largely focused on the development of stand-alone ML atmospheric and marine models \citep{Finn2024, aouni2025oceanbench, Cui2025, Dheeshjith2025, Guo2025, Huang2025, Gregory2026}, potentially forced by or coupled to data-driven or numerical atmospheric forecasts \citep{Wang2024,Aouni2025}. 
Our approach instead adopts a holistic training strategy across Earth system components, where fields from different components are used within a single model. 
The model\footnote{Our model is implemented within the Anemoi (\url{https://github.com/ecmwf/anemoi}) software framework, developed collaboratively by ECMWF and several European national meteorological services.} uses a single ML architecture with a shared latent representation. 
It simultaneously predicts multiple Earth system components: the atmosphere, surface ocean, sea ice, and surface waves.
We refer to this approach hereafter as joint modelling. 
In doing so, we build upon the extension of operational AIFS Single to land variables \citep{Moldovan2025}, as well as earlier joint-model efforts in the marine domain, most notably the Aurora model \citep{Bodnar2025}, which features a unified representation of the atmosphere and surface ocean waves.

A fundamental requirement for training a joint model is a consistent representation across components within the training datasets. 
As described in Sec. \ref{sec:datasets}, we address this requirement by training on observation-informed reanalyses produced at the European Centre for Medium-range Weather Forecasts (ECMWF). 
These datasets provide a highly consistent representation of past atmospheric and ocean waves conditions (ERA5; \citet{Hersbach2020}) and marine conditions (ORAS6; \citet{Zuo2024}).
A key potential advantage of the joint modelling approach is that it removes the need to impose a priori assumptions about information flow between components. 
Instead, the model can exploit the expressive capacity of modern ML architectures to learn cross-component relationships directly from the data. 
In fact, the model does not explicitly distinguish between Earth system components (e.g., atmosphere, ocean, sea ice, or waves). 
Instead, it treats all variables as part of a unified state space, making it effectively component-agnostic.
This is exemplified in Fig.~\ref{fig:tc_AIFS_IFS}, where tropical cyclone dynamics clearly impact all Earth system components represented in the joint ML model (AIFS Marine) and numerical model (IFS) alike. 

Using this joint modelling framework, we demonstrate high forecast skill for the newly introduced marine components.
For most surface ocean, sea ice, and wave variables, this corresponds to an improvement of roughly one day in medium-range lead time relative to the state-of-the-art ECMWF physics-based systems.
In addition, the joint models exhibit stable behaviour under strong perturbations of the initial conditions and respond in a physically consistent manner, supporting their robustness and applicability.

\section{Model Design and Training}

The AIFS architecture follows an encoder–processor–decoder structure. 
The encoder and decoder are both implemented as attention-based graph neural networks. 
The encoder maps input atmospheric and marine fields from the data grid into a latent representation, while the decoder projects the processed latent features back to the physical output grid. 
Between these components, the processor operates on the latent space using a transformer architecture with sliding-window attention (see \citet{Lang2024a} for further details).

Input variables for both the atmosphere and ocean are defined on an N320 reduced Gaussian grid, corresponding to an approximate horizontal resolution of 0.25°, consistent with the operational AIFS configuration. 
In the processor, computations are carried out on a latent grid based on an O96 octahedral reduced Gaussian grid \citep{Wedi2014} at about 1° resolution. 
The processor includes 16 layers, with attention windows structured along latitude bands.

The same encoder and decoder submodules are shared across atmospheric and marine components, reflecting the objective of producing forecasts for both systems at a common spatial and temporal resolution in a component-agnostic way. 
The joint AIFS model is trained to generate 6-hour forecasts ($t_0 + 6\mathrm{h}$) from the previous ($t_0 - 6\mathrm{h}$) and current ($t_0$) states. 
Longer lead times are obtained auto-regressively by time-stepping and iteratively feeding predictions back into the model, a procedure commonly referred to as rollout.

The implementation of the model and associated data pipelines builds on the Anemoi software ecosystem, developed by ECMWF in collaboration with several National Meteorological Services across Europe, which provides end-to-end support for dataset preparation, training, and inference.

\subsection{Datasets} \label{sec:datasets} 

In our work, we follow the consolidated approach for developing AIFS concerning the variable choice and vertical discretisation on 13 pressure levels \citep{Moldovan2025}.
The variables describing the atmospheric state are taken from ERA5 \citep{Hersbach2020}. 

To train the additional wave-related fields, a dedicated hindcast dataset covering the period 1979–2025 was produced using ECMWF’s most recent wave model (ecWAM) and altimeter wave height Data Assimilation (DA) system \citep{Bidlot2026prep} at a resolution of approximately 9 km. 
This hindcast includes a physically realistic representation of wave attenuation in the presence of sea ice \citep{Yu2022}.
This represents a major upgrade relative to earlier ECMWF products. 
In those, waves propagated through the sea ice unimpeded for sea-ice concentrations lower than 30\%, 
while for higher concentrations, the wave spectrum was reset to noise level and the wave output was masked.

For the surface ocean and sea-ice variables, we use hourly-averaged fields from the ECMWF Ocean Reanalysis System 6 (ORAS6) \citep{Zuo2024}. 
ORAS6 is based on the Nucleus for European Modelling of the Ocean version 4 \citep{Madec2024}, the SI3 sea ice model \citep{Vancoppenolle2023}, and the NEMOVAR ensemble three-dimensional variational data assimilation system \citep{Mogensen2012, Chrust2024}. 
The reanalysis is produced as a continuous integration spanning the period from 1993 to 2023, forced by hourly varying ERA5 atmospheric conditions, preceded by a five-year spin-up with active data assimilation from a cycled ocean state. 
The reanalysis assimilates in-situ temperature and salinity observations, sea-level anomaly (via satellite altimetry), sea-surface temperature, and sea-ice concentration. 
Observational constraints in the sub-surface ocean differ substantially before and after the introduction of Argo floats. 
The Argo programme was gradually deployed starting around 1999, providing increasingly routine and widespread monitoring of sub-surface temperature and salinity over the following years. 
Before then, sub-surface observations from ships and other platforms were much sparser in space and time. As a result, the three-dimensional circulation was constrained more strongly by satellite altimetry.
As this study focuses on surface variables, these regimes are not treated separately during training. 
ORAS6 provides one control integration and ten ensemble members. 
Here, we use only the control member, leaving explicit exploitation of ensemble information to future work. 
Although only surface ocean and sea-ice variables are used for training, ORAS6 is a fully three-dimensional reanalysis with explicit sea ice representation. 
The reanalysis uses an ORCA025 grid, which corresponds to a resolution of approximately 25 km. 

Both marine training datasets—the wave hindcast and ORAS6—are forced by ERA5, ensuring consistency when jointly training the model across different Earth system components. 
The ERA5 forcings used for ORAS6 include many wave-related processes, such as wave-induced Stokes drift, wave-breaking turbulent kinetic energy (TKE) surface flux, and wave-modulated surface stress forcing from the wave-model analysis. 
These processes provide an additional momentum pathway between surface waves and the ocean and influence near-surface currents and upper-ocean mixing.

For the fine-tuning stage of the training (see section \ref{sec:trainingschedule} for details), additional marine training datasets are generated for model fine-tuning and rollout training.
To this end, we force ORAS6 and the wave hindcast with the ECMWF operational analysis, rather than the ERA5 reanalysis, for the necessary training years.

\subsection{Variable Selection} 

For the wave component, we deliberately reduce the dimensionality relative to traditional physics-based wave models, as is also done for the atmosphere, where 137 model levels from IFS are reduced to 13 pressure levels in the AIFS \citep{Lang2024a}.
In ecWAM, the full two-dimensional wave spectrum is explicitly discretised in frequency and direction. 
In the current deterministic forecast configuration, this corresponds to 36 frequency bins and 36 directional bins, yielding over 1,200 spectral components per grid point \citep{ECMWF2024IFS}. 
This detailed spectral representation allows the model to resolve the full distribution of wind-seas and swells, but it comes with high computational cost and data dimensionality.
In our joint ML model, we instead adopt a reduced set of variables that captures the essential characteristics of the wave spectrum. 
Specifically, we include significant wave height, mean wave period, mean wave direction, and the wave-dependent drag coefficient\footnote{Wave-related drag coefficient connects winds to the total momentum transfer from atmosphere to ocean, including wave effects.}. Together, these variables represent integral measures of the wave state and its coupling to the atmosphere. 
To retain information about the wave spectral distribution and to provide useful predictions for marine stakeholders, 
we further model the decomposition of significant wave height into six distinct period bands for all waves with periods larger than 10 seconds.
These variables are also physics-based model output and allow the ML model to differentiate between locally generated wind-seas and remotely generated swell systems. 
This balance provides physical expressiveness and keeps the number of variables representing the wave component in the model tractable within a coupled Earth system setting.

For the ocean, the ML model uses a reduced set of variables compared to the numerical ocean model from which the training data are derived. 
Only surface variables are considered. 
In particular, we use sea surface temperature and salinity, which are prognostic variables in the numerical model; sea surface height, which implicitly contains information about the barotropic flow of the ocean; and the zonal and meridional components of the surface currents.

We adopt a reduced-dimensional representation of the sea ice component as well. 
In the numerical model SI3, sea ice is described using a subgrid-scale discretisation into five thickness classes \citep{Lipscomb2001, Massonnet2019}. 
Within each thickness class, sea ice and overlying snow thermodynamics are further discretised into multiple vertical layers \citep{Vancoppenolle2009}. 
In the proposed ML counterpart, this complexity is reduced to a set of five grid-cell-averaged prognostic variables: sea ice concentration, sea ice albedo, sea ice volume per unit area, snow volume over sea ice per unit area, and the zonal and meridional components of the sea ice velocity. 

This simplification is justified for two reasons. First, ML models can represent nonlinear effects associated with subgrid-scale processes without resolving them explicitly. Second, atmospheric applications have shown that such models can operate on reduced representations while retaining essential dynamics (e.g., reduced vertical discretisation in AIFS compared to IFS \citep{Lang2024a}). Additionally, although the data assimilation system does not operate natively on the full multi-category ice thickness distribution, it assimilates sea ice concentration through a single-category control vector, with increments subsequently distributed across categories in a manner consistent with the existing ice state, leading to a coherent temporal evolution of sea ice \citep{Browne2025}. 

To allow the marine fields to be easily combined with the variables from the atmosphere, they are linearly interpolated onto the N320 grid. 
A complete list of all model variables considered in this study, together with their characteristics, is provided in Table~\ref{tab:variables}.

\subsection{AIFS Model Versions}

We train and evaluate four AIFS model variants that differ in the Earth system components they represent. 
The models follow the same core design, with differences arising from the inclusion of oceanic and/or wave dynamics and corresponding adjustments in model capacity. 
For configurations that include the ocean component, the number of channels in the processor is increased by 50\% to account for the higher complexity of coupled atmosphere--ocean interactions. 
These variants are designed to isolate the impact of additional Earth system components on forecast skill while maintaining a consistent architectural baseline. 
An overview of the resulting models and their parameter counts is provided in Table~\ref{tab:trainable_param}.

\begin{table}[ht]
    \centering
    \begin{tabular}{lcccc}
    Model &
    Components &
    \makecell{Channels in\\Processor} &
    \makecell{Trainable\\Parameters} \\ \hline
    AIFS Atmosphere & Atmosphere &  1024 & 253 M \\
    AIFS Waves & \makecell{Atmosphere \&  Waves} & 1024 & 253 M \\
    AIFS Ocean & \makecell{Atmosphere \&  Ocean \& Sea Ice} & 1536 & 539 M \\
    AIFS Marine & \makecell{Atmosphere \& Ocean \& Sea Ice \& Waves} & 1536 & 539 M
    \end{tabular}
    \caption{Model versions with corresponding number of channels in processor and total number of trainable parameters.}
    \label{tab:trainable_param}
\end{table}

Details of the IFS forecast configurations used as baselines for comparison with the ML models are provided in Sec.~\ref{sec:ifs} of the supplementary material.

\subsection{Training Schedule}
\label{sec:trainingschedule}

The training of the different joint AIFS models detailed above follows a two-stage procedure, which is also used in the operational AIFS \citep{Moldovan2025}. 
In the pre-training phase, the model is trained to predict the atmospheric state 6 hours ahead, providing a robust initialisation of the model parameters. 
In previous work \citep{Lang2024a,Moldovan2025}, AIFS is pre-trained on ERA5 data from 1979 onward.
However, as the ORAS6 reanalysis is only available from 1993 at the moment, all joint model variants are trained over the period 1993–2022.
This restriction has a negligible impact on forecast performance, as shown in Sec.~\ref{sec:effect_training_years} in the supplementary material.

In the second rollout fine-tuning phase, the model is trained auto-regressively for lead times up to 72 hours (12 time steps) on the years 2016 to 2022. 
This design allows the model to learn from its own predictions. 
Fine-tuning the model on recent years also aligns it with the current climate regime and benefits from higher-quality and denser observations.

Further details on learning rate schedules, optimiser settings, and rollout progression follow those of AIFS v1.1 \citep{Moldovan2025}.

\section{Technical Development}

Our joint modelling approach for ocean waves, sea ice, and the surface ocean required several technical adaptations of the AIFS \citep{Lang2024a,Moldovan2025} to ensure physical consistency, numerical stability, and balanced training across components. In the following paragraphs, we summarise these developments.

\begin{figure}
    \centering
    \adjustbox{center}{%
        \includegraphics[width=1.0\linewidth]{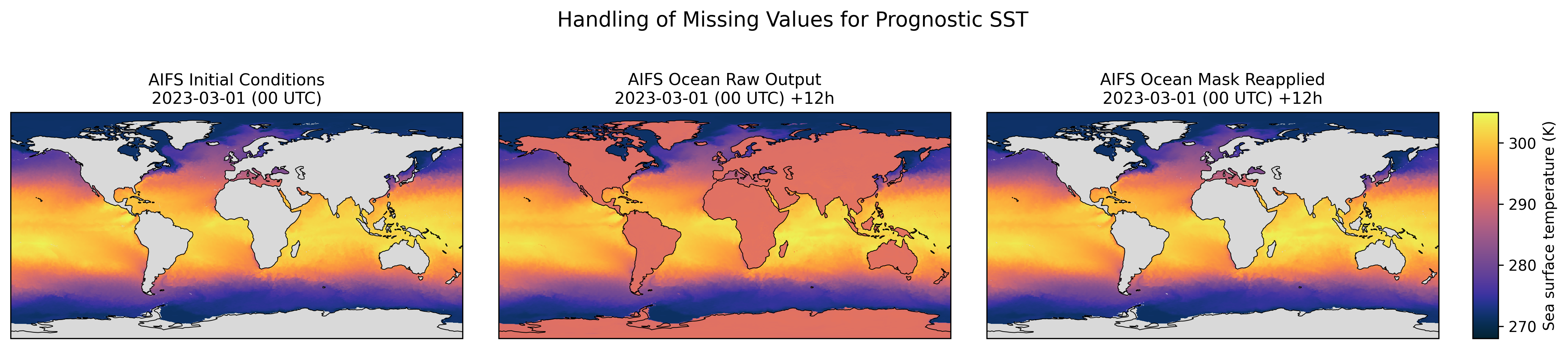}
    }
    \caption{
    Illustration of the handling of missing values for prognostic sea surface temperature (SST).
    The input field (01.03.2023, 0 UTC) has missing values over land (left). 
    Because missing values over land are replaced with zeroes in normalised space during preprocessing, the raw model output (12 h forecast from AIFS Ocean) at these locations remains close to the background state, as no meaningful increment is learned there (middle). 
    In the final output after reapplying the missing-value mask, the missing values are restored at grid points where the variable is physically undefined (right).
    }
    
    \label{fig:nan_handling}
\end{figure}

\subsection{Handling Missing Data}
Several variables are not defined on the full model domain, which leads to missing values (\texttt{NaN}s) in the input data. 
For example, all newly added marine fields contain missing values over land points.  
Because standard ML models cannot process \texttt{NaN}s directly, we replace them with zeroes in normalised space before the data is fed into the network to ensure stable training and forecasting. 
At the same time, we retain a mask that records where the values are undefined. 
During training, this mask is used to exclude undefined locations from the loss function, so that the model is not penalised for predictions in regions where the variable is not physically defined.
During inference, the mask is applied again to restore missing values at the appropriate grid points. 
This ensures that the final outputs remain consistent with the physical domain of each variable. 
Figure~\ref{fig:nan_handling} illustrates this procedure for a representative marine field.

\begin{figure}
    \centering
    \includegraphics[width=0.35\linewidth,clip,trim=0 296 0 18]{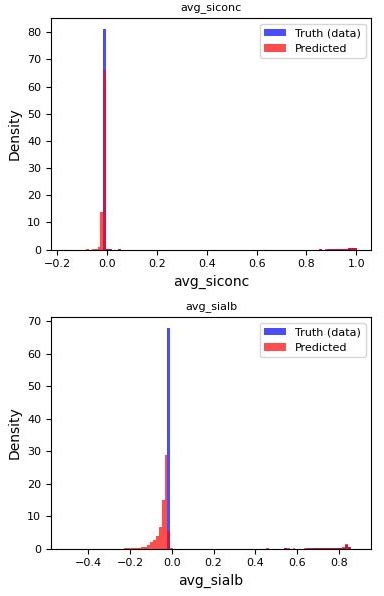}
    \includegraphics[width=0.35\linewidth,clip,trim=0 297 0 18]{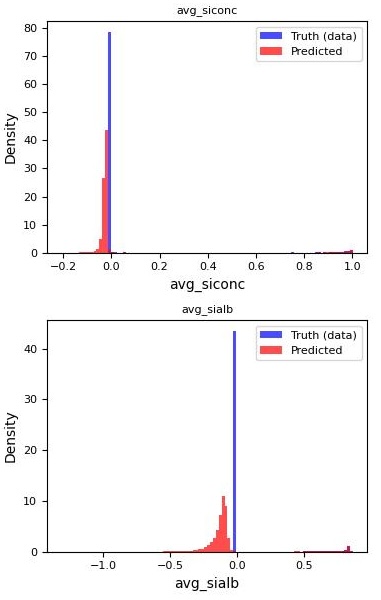}
    \caption{Histogram of 6h sea ice concentration predictions by the model AIFS Ocean before applying the bounding function. 
    The outputs of a model trained with LeakyHardTanh bounding are more concentrated near the physical lower bound 0 (left) when comparing to a model trained with HardTanh bounding (right).}
    \label{fig:leakybounding}
\end{figure}

\subsection{Physical Consistency and Bounding}

To guarantee physically meaningful outputs for the newly modelled variables, we enforce variable-specific bounds, similar to those enforced by the physics-based models SI3 and ecWAM. 
To this end, we extend the bounding that has already been applied to precipitation fields in \citep{Moldovan2025}.
Following these results, we apply a ReLU bounding function to non-negative wave variables during both training and inference. 

For sea-ice variables, values are exactly at the lower physical bound (zero ice) over large ocean regions. To avoid vanishing gradients in these regions, we use a leaky ReLU with $\alpha = 0.01$. 
\begin{equation*}
\mathrm{ReLU}(x) = \max(0, x) 
\qquad  \qquad
\mathrm{LeakyReLU}(x) = 
\begin{cases}
x, & x \geq 0 \\
\alpha x, & x < 0
\end{cases}
\text{ \hspace{2ex} with } \alpha \in (0,1)
\end{equation*}
This allows small weight updates even when predictions fall in the negative domain.
Leaky ReLU bounding functions are also applied to sea surface salinity and temperature fields. 
For sea surface temperature, a lower threshold of 271.15\,K is imposed through the leaky formulation to discourage unphysical values, as temperatures below the seawater freezing point are not physically consistent. 
In its present form, this bounding threshold does not account for the seawater freezing temperature dependence on salinity.

For sea ice fields describing concentrations between 0 and 1, we apply leaky Hardtanh-based bounding functions 
\begin{equation*}
\mathrm{HardTanh}_{0,1}(x) = \min(1,\max(0, x))
\end{equation*}
\begin{equation*}
\mathrm{LeakyHardTanh}_{0,1}(x) = 
\begin{cases}
x, & x \in [0,1]\\
\alpha x, & \text{otherwise}
\end{cases}
\text{ \hspace{2ex} with } \alpha \in (0,1).
\end{equation*}

An analysis of the model outputs prior to applying the bounding function shows that leaky bounding produces predictions that are consistently closer to the physical lower bound (see Fig.~\ref{fig:leakybounding}).
In contrast, standard (non-leaky) bounding maps all negative values to zero, leading to vanishing gradients in these regions and thereby limiting further correction during training.
The leaky formulation avoids this issue by preserving small negative values, which remain visible to the loss function and can therefore be penalised.
This enables continuous weight updates near the boundary and encourages the model to place predictions closer to the physically admissible range.
The effect is particularly pronounced for sea ice variables, where the target values are exactly zero over large ocean regions.

Additional post-processing steps are applied during inference and do not affect gradients during training.
In particular, forecast fields that were bounded by leaky ReLU or leaky HardTanh functions during training are mapped into their valid physical ranges using a non-leaky formulation. 

We impose additional consistency constraints during post-processing in inference on the forecasted sea ice fields.
When sea ice concentration is zero, all other sea ice variables are set to zero. 
This constraint prevents physically implausible outputs in ice-free regions and reduces error accumulation in long-range autoregressive forecasts, for example, in the sea ice velocity fields. 

Table \ref{tab:variables} in the supplementary material lists the constraints applied to the variables to ensure physical consistency in the AIFS model versions.

\subsection{Directional Variable Remapping}
To improve forecasts of the directional variable mean wave direction (MWD, in degrees), we remap the angular quantities into their cosine and sine components
\begin{equation*}
\text{sin\_mwd} = \sin(\text{MWD}), \quad 
\text{cos\_mwd} = \cos(\text{MWD}),
\end{equation*}
similar to the treatment of temporal variables in the AIFS \citep{Lang2024a}. This avoids discontinuities between 0° and 360° and provides a smooth representation for the model. 

\subsection{Loss Scaling of Variables}
We apply dedicated loss scaling strategies to stabilise training across heterogeneous variables. 

First, scaling factors are introduced to account for differences in dynamical timescales, creating a more uniform optimisation landscape. 
Many surface ocean and sea ice variables evolve more slowly than atmospheric fields. 
Since the model predicts increments via a skip connection, slowly varying fields contribute less strongly to the loss. 
To compensate for this, these variables, including most surface ocean and sea ice fields, are assigned larger scaling factors so that their temporal evolution is adequately captured by the model.  

Second, we weight the contributions of new wave, ocean, and sea ice variables in the loss function to balance their influence relative to atmospheric components. 
This ensures that performance improvements in one subsystem do not come at the expense of degradation in others.
The loss scaling factors were determined empirically and are listed in Table \ref{tab:variables} for each variable and model configuration (AIFS Atmosphere, AIFS Waves, and AIFS Ocean). 

To maintain atmospheric forecast quality when training the full AIFS Marine model—including both the surface ocean and ocean waves—it was necessary to reduce the loss weights of most marine variables by a factor of two.
The scaling factors for the atmospheric fields, including the weighting of different pressure levels, are taken from \cite{Moldovan2025}.

\section{Results}

In this section, we evaluate the performance of the joint AIFS configurations across the newly introduced marine components - surface ocean, ocean waves, and sea ice - and assess their impact on atmospheric forecast skill. 
We first present quantitative verification results for waves, sea ice, and the surface ocean, and then analyse how these components influence atmospheric forecasts. 
Finally, we examine coupled case studies and sensitivity experiments to assess the physical consistency and robustness of the joint modelling framework.

\begin{figure}
    \centering
    \raisebox{0.33cm}{%
        \includegraphics[width=0.35\linewidth]{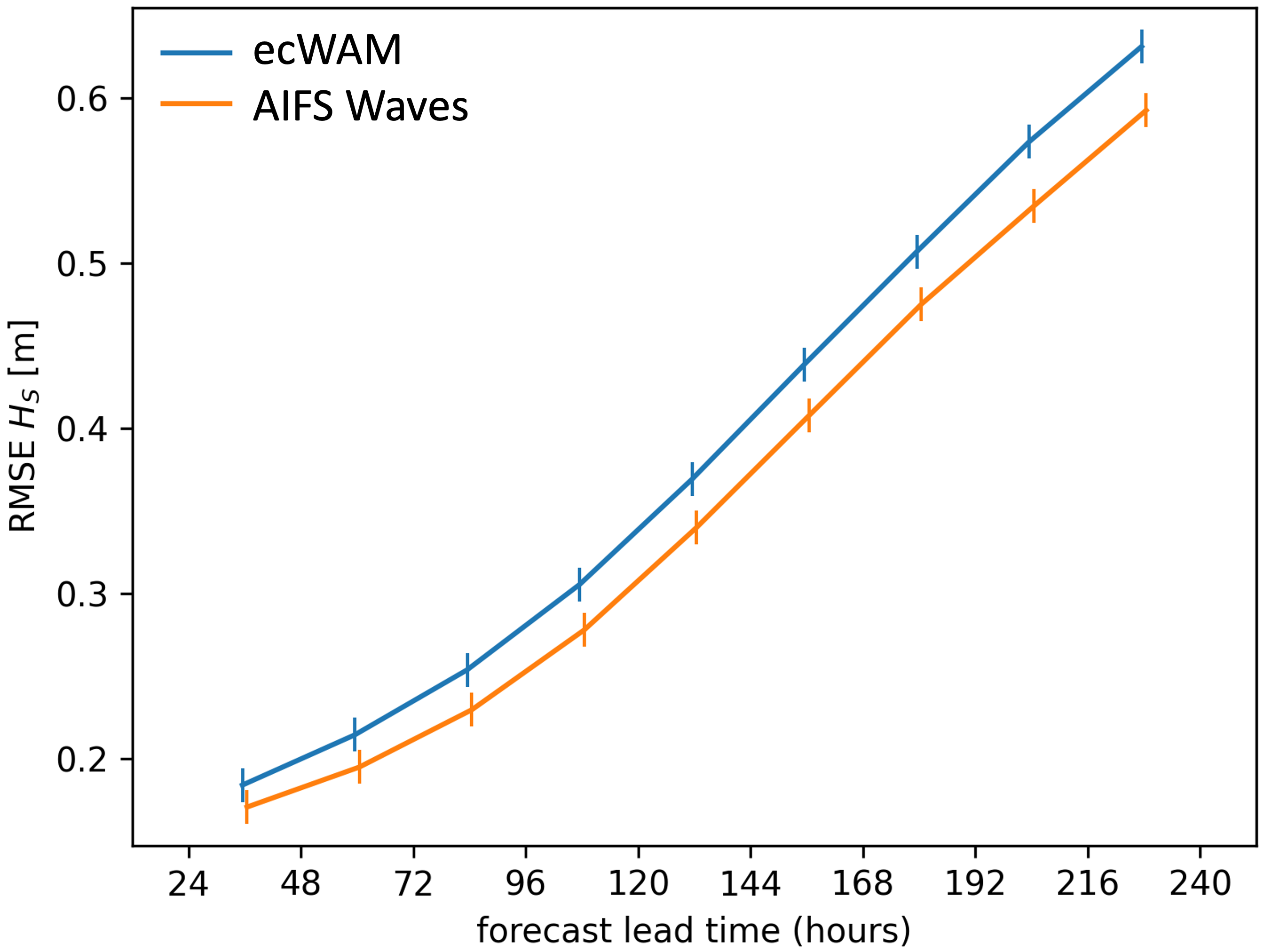}

    }
    \includegraphics[width=0.5\linewidth,clip,trim=0 5 0 5]{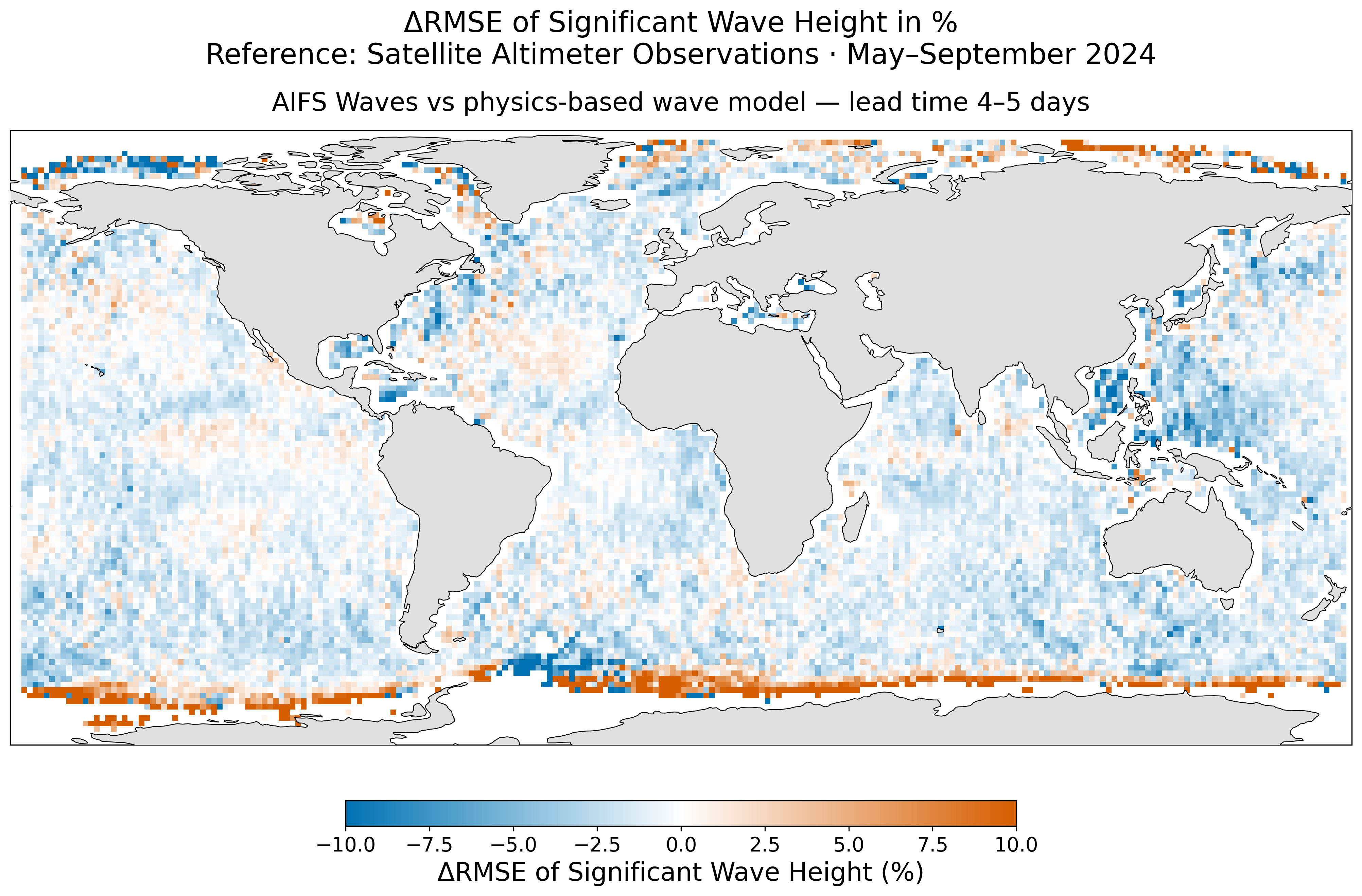}
    \caption{    
    Comparison of AIFS Waves and the physics-based wave model in root mean square error (RMSE) of significant wave height forecasts against satellite altimeter observations for May–August 2024. 
    RMSE over time for the joint atmosphere–wave model prototype (orange) and the physics-based baseline (blue), where lower values indicate better performance (left). 
    Global map of RMSE differences for lead time of 4 to 5 days, where blue indicates an improvement for the ML model and red a degradation (right).}
    \label{fig:scores_waves}
\end{figure}

\begin{figure}
    \centering
    \includegraphics[width=0.98\linewidth,clip,trim=0 5 0 5]{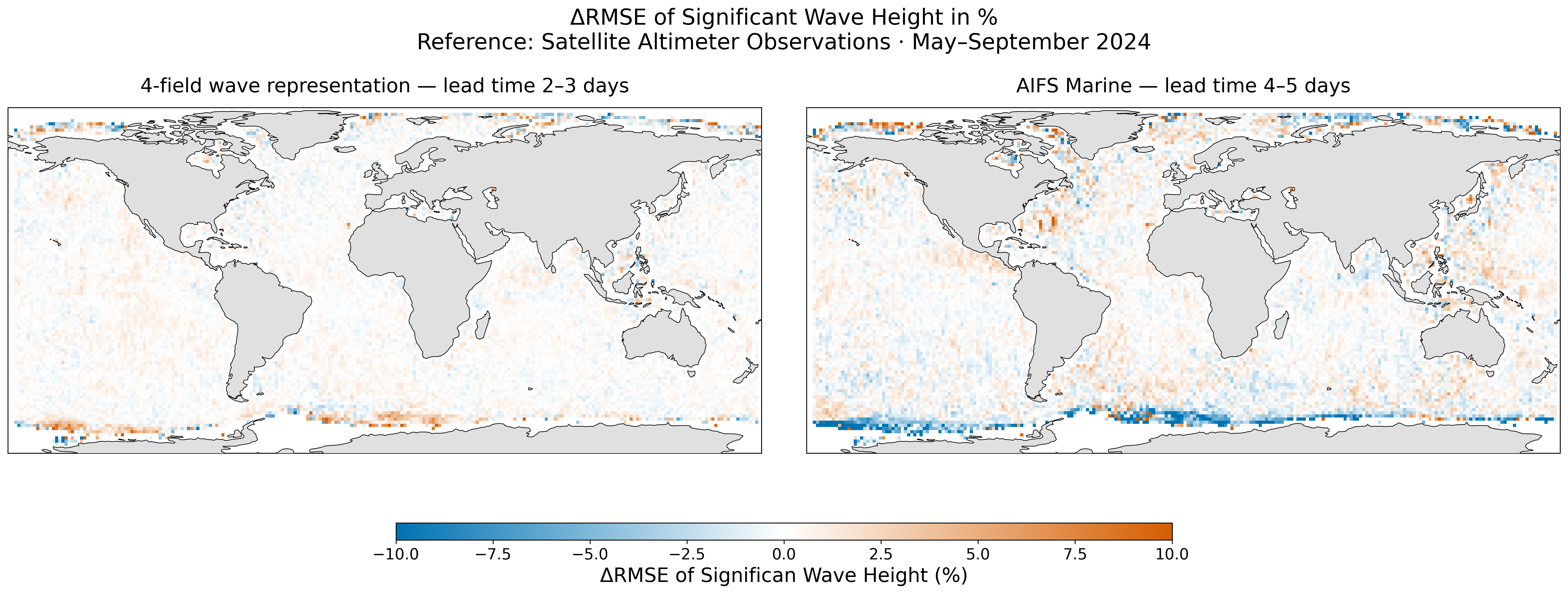}
    \caption{Comparison of two variations of a joint model to AIFS Waves in RMSE of significant wave height (SWH) forecasts against satellite observations, May–August 2024. Blue indicates an improvement in RMSE, and red indicates degradation. 
    Using a model with a smaller 4-field wave representation degrades model performance along the sea ice edge in comparison to AIFS Waves (lead time of 2 to 3 days, left). 
    The explicit sea ice information in AIFS Marine noticeably improves SWH forecasts along the sea ice edge in comparison to AIFS Waves (lead time of 4 to 5 days, right).}
    \label{fig:diff_rmse_more_waves}
\end{figure}

\subsection{Waves}

We evaluate the performance of different AIFS model configurations in forecasting ocean wave conditions, with a primary focus on significant wave height forecasts. 

Incorporating wave variables into AIFS leads to clear improvements in the representation and forecasting of ocean wave conditions relative to the physics-based baseline model. 
When validated against buoy and satellite observations (see Fig.~\ref{fig:scores_waves_buoys} and Fig.~\ref{fig:scores_waves} respectively), the data-driven model AIFS Waves reduces the medium-range forecast error for SWH by approximately 10 \% compared to ECMWF’s operational wave forecasting system.
This corresponds to an improvement of roughly one day in lead time, with gains observed across most regions globally.

To further assess the representation of extremes, we analyse the distribution of forecasted variables against observations using quantile–quantile diagnostics, see Figure \ref{fig:wave_extremes}. 
For near-surface wind speed, we confirm a known limitation for MSE-trained data-driven models: AIFS Waves tends to underrepresent the strongest winds \citep{Lang2024a,Bouallegue2023}. 
In contrast, this behaviour is not reflected in the significant wave height forecasts. The wave hindcast dataset used for training includes a statistical correction of wind forcing, leading to a more accurate representation of extreme wave conditions. 
In addition, significant wave height fields exhibit fewer fine-scale features than surface wind fields. 
As a result, the underestimation of atmospheric extremes does not directly translate into negative biases in the upper tail of significant wave height distributions.

To better understand the impact of variable choice, we compare two configurations: one including only four prognostic wave variables (significant wave height, mean wave period, mean wave direction, and drag coefficient) and AIFS Waves with ten wave-related fields (additional decomposition of significant wave height into six distinct period bands). 
The extended configuration yields slightly improved accuracy, particularly along the sea ice edge, where additional spectral information helps capture complex wave–ice interactions (see Fig.~\ref{fig:diff_rmse_more_waves}, left).
Beyond improving skill, frequency-resolved wave information enhances forecast utility, as long-period swell often dominates coastal and offshore hazards due to its higher energy and impact potential.

Despite these improvements, the AIFS Waves exhibits reduced skill near the sea ice edge, where SWH is rapidly attenuated to near-zero values, when compared to the physics-based baseline. 
As expected for root mean square error (RMSE)-minimising models, this sharp transition is difficult to learn and partially smoothed. 
This smoothing behaviour is more pronounced in AIFS Waves, whereas AIFS Marine shows improved performance in these regions due to the explicit representation of sea ice (Fig.~\ref{fig:diff_rmse_more_waves}, right). 
This difference is further illustrated in Fig.~\ref{fig:sea_ice_edge_waves}, where the sea ice edge appears more sharply defined in AIFS Marine forecasts.

Spatial analyses further demonstrate that the joint atmosphere–wave model captures key physical features. 
Figure~\ref{fig:waves_islands} shows the shadowing effect of islands, including those not resolved at the model grid scale. 
In addition, Sec.~\ref{sec:coupled_test_cases} highlights consistent interactions across components, such as wave damping near sea ice and the alignment between wave and surface wind fields during tropical cyclones.

\begin{figure}
    \centering
    \includegraphics[width=0.85\linewidth]{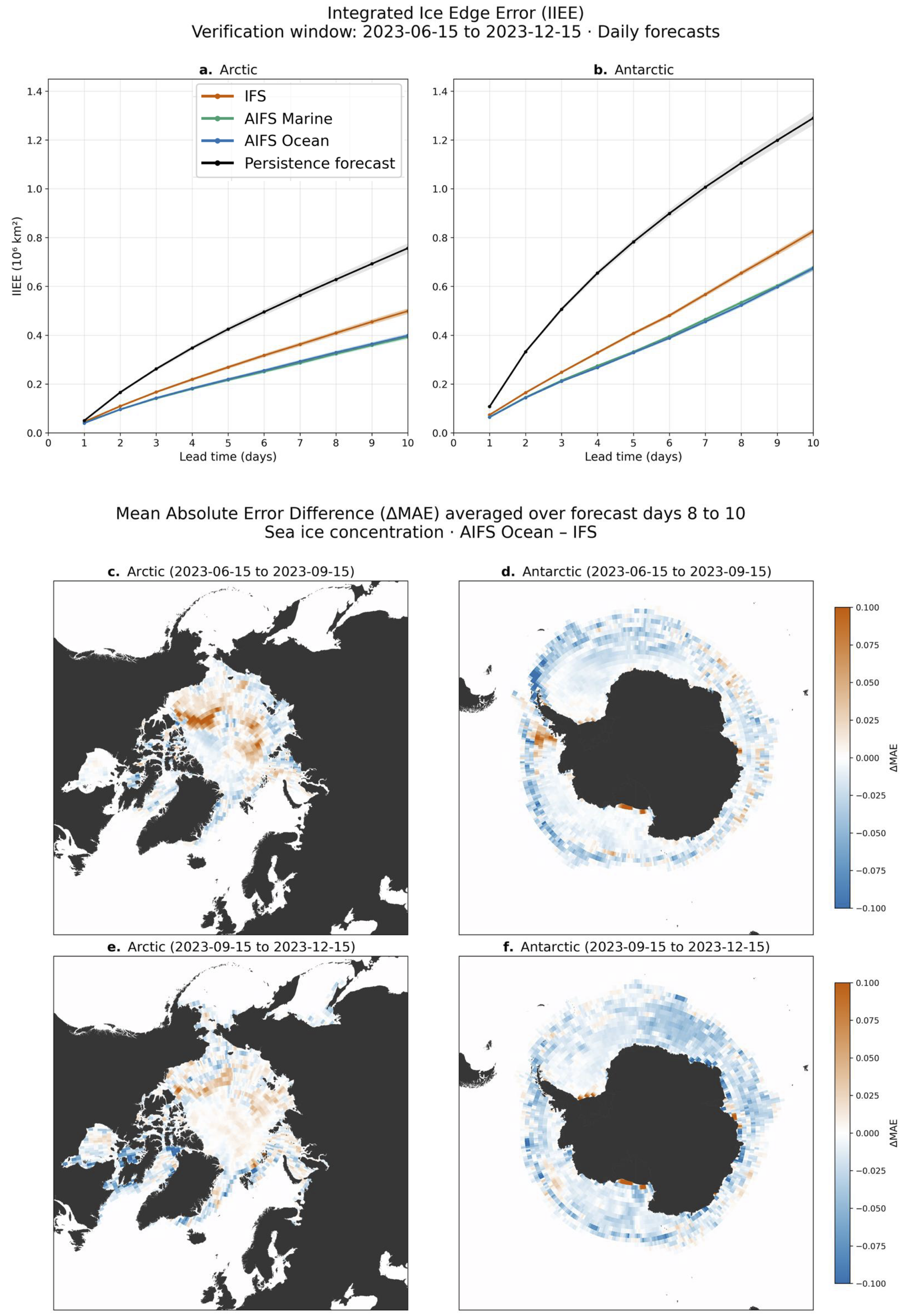}
    \caption{
    (\textbf{a,b}) Integrated Ice Edge Error (IIEE) for the Arctic and Antarctic as a function of lead time, verified against ORAS6 for 15 June--15 December 2023.
    (\textbf{c--f}) Spatial maps of the Mean Absolute Error difference ($\Delta$MAE) in sea ice concentration between AIFS Ocean and IFS, averaged over forecast days 8--10, for two sub-periods: 15 June--15 September 2023 and 15 September--15 December 2023.
    }
    \label{fig:sea_ice_scores}
\end{figure}

\subsection{Sea Ice}

Fig.~\ref{fig:sea_ice_scores} illustrates both the overall skill of the different systems in predicting the sea ice edge and the spatial structure of their concentration errors. The top panels show the Integrated Ice Edge Error (IIEE; \citealp{Goessling2016}) as a function of lead time, verified against ORAS6. The IIEE is a widely used metric (e.g., \citealp{Zampieri2018,Bushuk2024}) and is particularly well suited to evaluating sea-ice edge position. It measures the total area where ice is either falsely predicted or missed, thereby combining information on both ice advance and retreat into a single physically interpretable scalar. It is therefore more sensitive to displacements of the ice edge than grid-point-based scores. For reference, the persistence forecast—shown in black—corresponds to a trivial prediction in which the initial sea ice field is kept constant in time, and provides a useful baseline. In fact, persisted sea-ice cover was used in NWP models, including IFS, before the transition to Earth system models. Finally, the metrics are computed after interpolating both the ML and physics-based fields onto a common regular grid with a resolution of 0.5$^\circ$.

Both AIFS configurations that include the ocean/sea-ice component (AIFS Ocean and AIFS Marine) substantially reduce IIEE relative to the IFS in both hemispheres, with particularly pronounced improvements in the Southern Ocean. 
Note that the IIEE curves for AIFS Ocean and AIFS Marine are almost indistinguishable, indicating that the explicit wave information does not bring additional benefit for defining the ice edge. 
This is not surprising since in the training data, the wave hindcast responds to the presence of sea ice, but the sea ice itself does not directly respond to wave forcing.
Therefore, waves do not provide an independent control on ice-edge evolution. The lower panels illustrate where these improvements occur by showing the mean absolute error difference in sea ice concentration ($\Delta$MAE; AIFS Ocean minus IFS) averaged over forecast days 8–10 for two sub-periods. The magnitude of $\Delta$MAE is generally small, reaching at most about 10\% of the sea ice concentration, but with coherent and widespread improvements (blue) along large parts of the marginal ice zone, especially in the Southern Ocean. These Southern Ocean improvements are particularly encouraging because the marine representation in the joint model is purely two-dimensional. It does not include explicit information about the three-dimensional temperature and salinity structure of the ocean, even though that structure can influence sea-ice evolution on these timescales. At the medium range, however, dynamics remains the primary driver of Southern Ocean sea-ice variability.

\subsection{Surface Ocean}

\begin{figure}
    \centering
    \adjustbox{center}{%
        \includegraphics[width=1.1\linewidth]{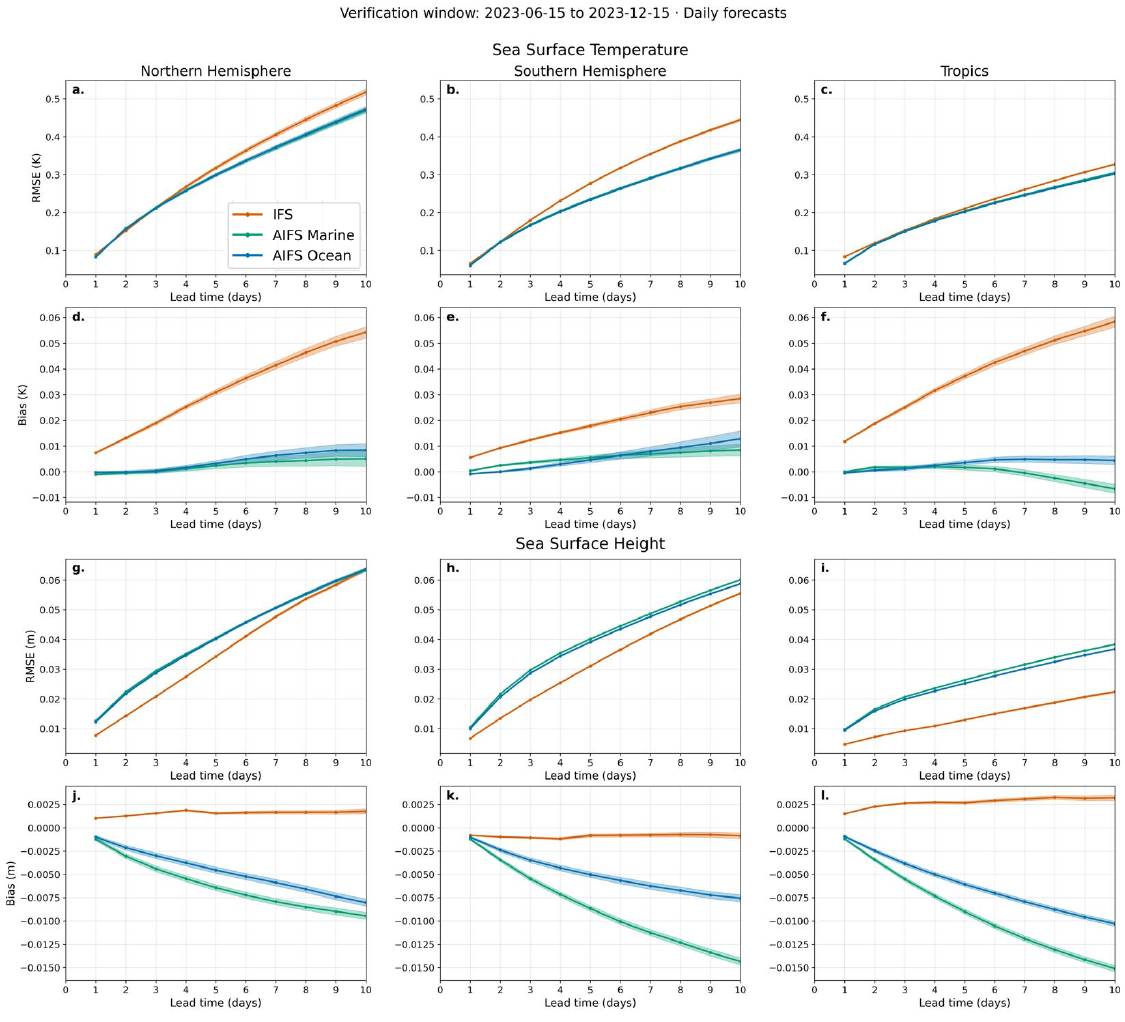}
    }
    \caption{
    Forecast verification of sea surface temperature (top two rows) and sea surface height (bottom two rows) against the reference analysis for the period 15 June--15 December 2023. Columns show results for the Northern Hemisphere, Southern Hemisphere, and Tropics. Rows show RMSE (first and third rows) and mean bias (second and fourth rows). Curves are shown for IFS, AIFS Marine, and AIFS Ocean as a function of forecast lead time.
    }
    
    \label{fig:ocean_scores}
\end{figure}

Fig.~\ref{fig:ocean_scores} summarises the forecast performance of the joint models for sea surface temperature (SST) and sea surface height (SSH), both verified against ORAS6, which—like sea ice concentration—is strongly constrained by the assimilation of satellite and in situ observations. The improvement of AIFS prediction skill over IFS differs clearly between these two variables. For SST, both AIFS Ocean and AIFS Marine show systematic improvements over the IFS across all regions, with lower RMSE and, in particular, substantially reduced biases in the Northern and Southern Hemispheres. The RMSE improvements are smaller in the tropics, where SST variability is weaker. In the tropics and northern hemisphere, we observe a substantial bias reduction compared to the physics-based system, while the reduction is moderate in the southern hemisphere.

For SSH, by contrast, the situation is reversed. While the RMSE of the ML models is comparable to or slightly worse than that of the IFS, both AIFS Ocean and AIFS Marine exhibit a systematic negative bias that increases with lead time in all regions, indicating a tendency to underestimate sea surface height. A plausible explanation is that SSH contains a persistent, nearly monotonic climate-change signal over the training period, namely global mean sea-level rise. As a result, the model may tend to predict a more climatological state that reflects the average conditions seen during training. In future versions of the system, this issue could likely be mitigated by training the model to predict detrended SSH anomalies rather than absolute values. Overall, these results highlight both the clear potential of the joint approach for surface temperature prediction and a key remaining challenge for representing slowly varying, trend-dominated ocean variables such as sea surface height.

\subsection{Impact on the Atmosphere}

\begin{figure}
    \centering
    \includegraphics[width=0.45\linewidth]{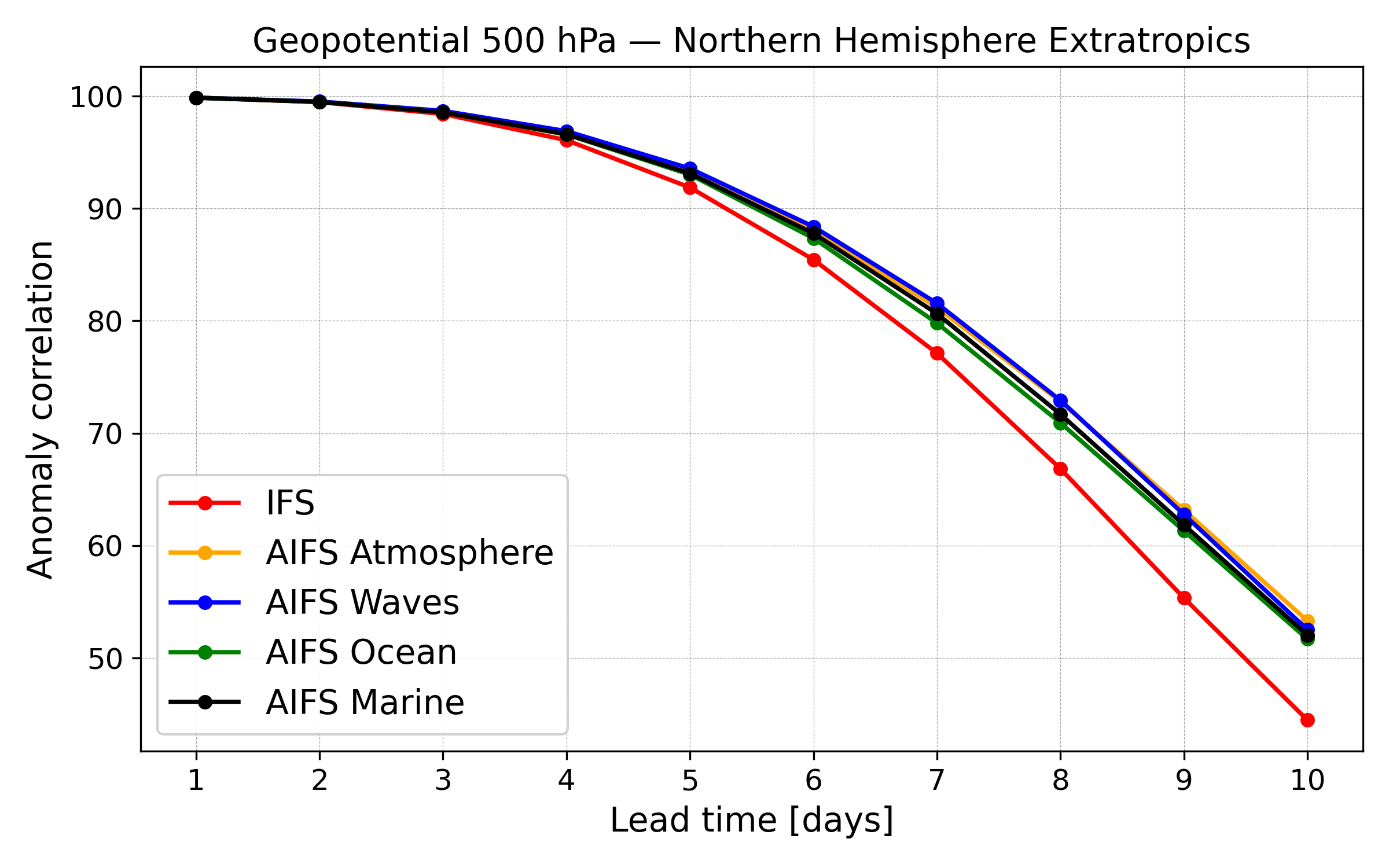}
    \includegraphics[width=0.45\linewidth]{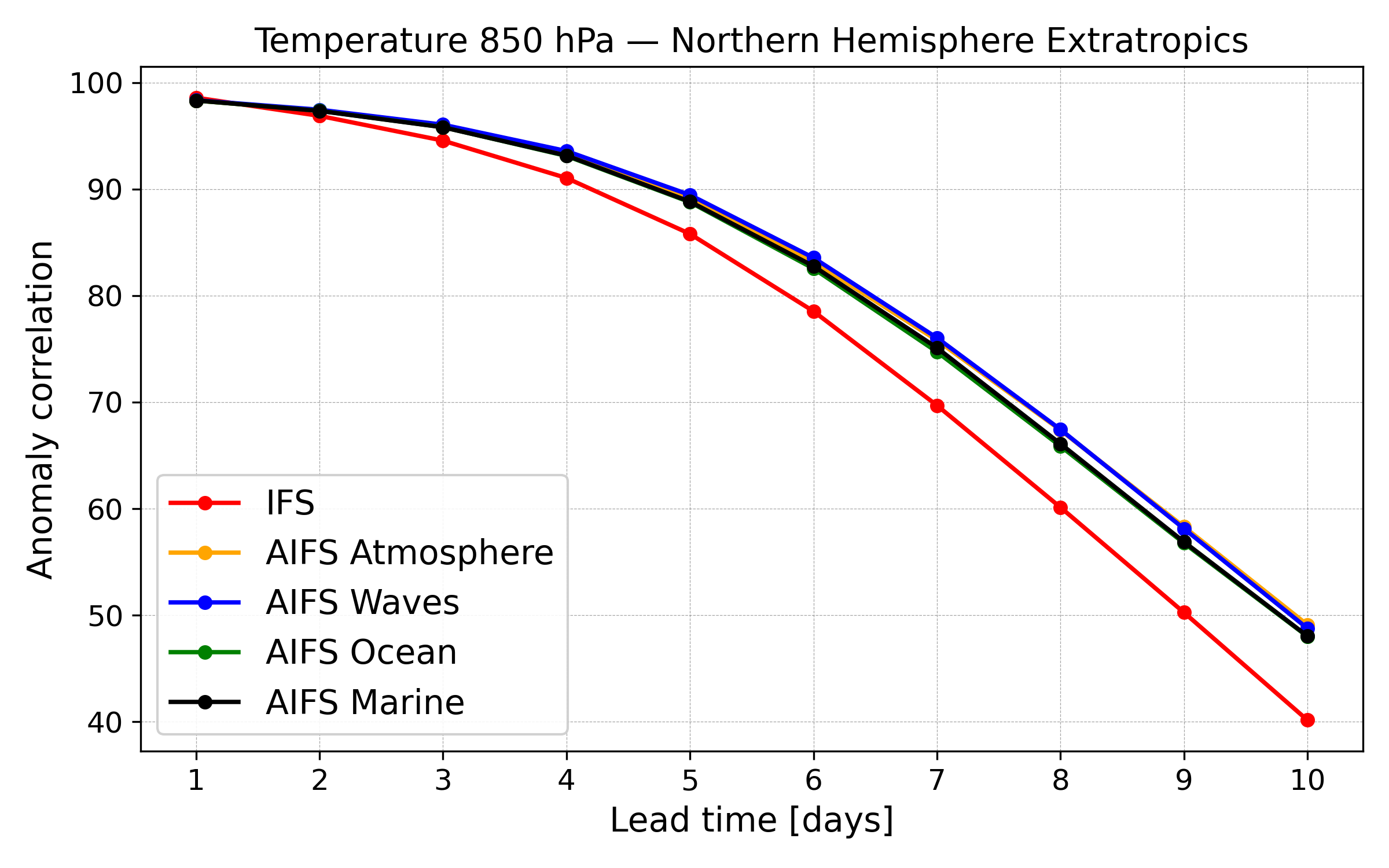}
    \caption{Anomaly correlation skill scores for geopotential at 500hPa (left) and temperature at 850hpa (right) in the Northern Hemisphere Extratropics. Skill scores computed for 15 June--15 December 2023 against IFS analysis.}
    \label{fig:upper_air_scores}
\end{figure}

\begin{figure}
    \centering
    \includegraphics[width=0.45\linewidth]{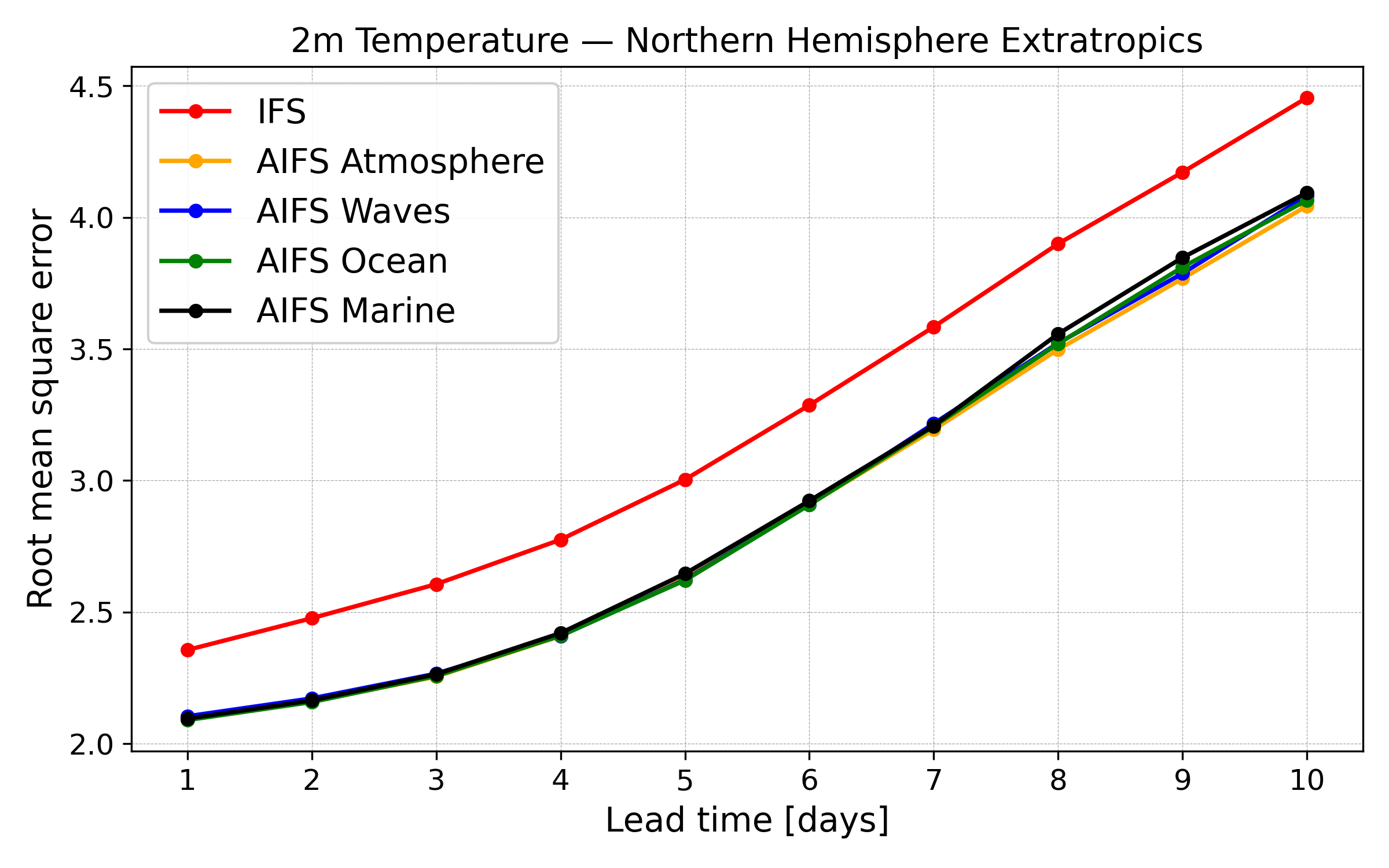}
    \includegraphics[width=0.45\linewidth]{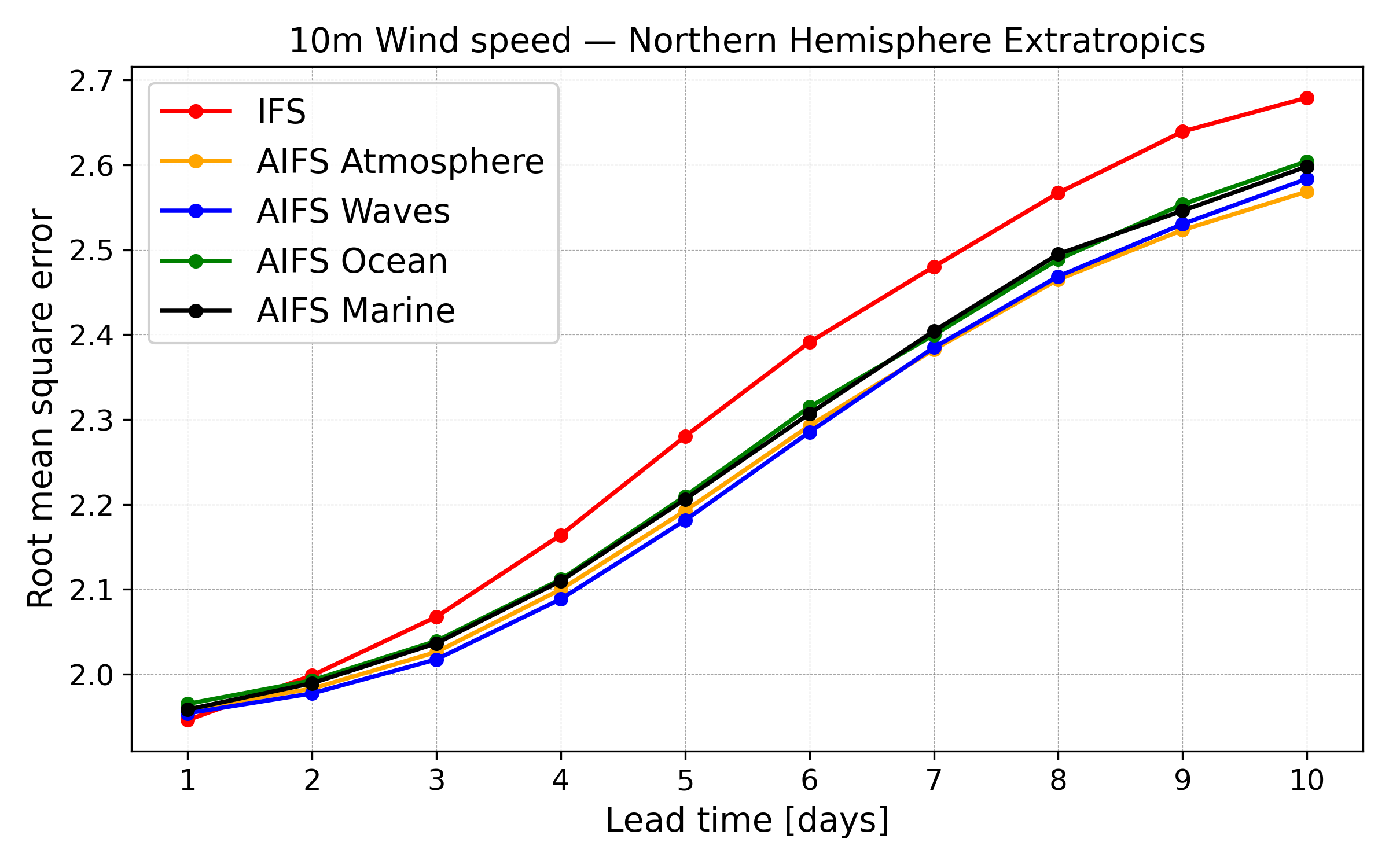}
    \caption{Root mean squared error (RMSE) scores for 2-metre temperature (left) and 10-metre wind speed (right) computed against SYNOP observations over the Northern Hemisphere for 15 June--15 December 2023.}
    \label{fig:sfc_obs}
\end{figure}

Overall, the inclusion of surface ocean waves in the AIFS Waves has a neutral impact on atmospheric forecast scores (Figs.~\ref{fig:upper_air_scores}, \ref{fig:sfc_obs}, and \ref{fig:scorecard_wave}). We attribute this to two main factors. 
First, the direct dynamical influence of waves on the atmosphere is limited at medium-range timescales. 
More importantly, we assume that the AIFS has already learned an intrinsic representation of the most relevant wave–atmosphere interactions through its training data. Both ERA5 and the operational IFS analysis explicitly represent surface waves, and their dynamical influence is therefore implicitly encoded in the atmospheric state provided to the model during training. As a consequence, adding an explicit wave component introduces limited additional information beyond what is already embedded in the dataset, resulting in only marginal changes in atmospheric skill.

The impact becomes more heterogeneous when explicit representations of the surface ocean and sea ice are added to the joint model. While the overall performance remains broadly comparable, we observe a slight degradation in selected atmospheric forecast scores, despite an increase in model capacity to account for the more complex prediction task, see Fig.~\ref{fig:upper_air_scores}. To maintain atmospheric forecast quality when training the full marine model—including both the surface ocean and ocean waves—it was necessary to reduce the loss weights of most marine variables by a factor of two. This trade-off suggests increased competition for representational capacity between atmospheric and marine components within the shared latent space.

This degradation highlights a fundamental challenge of joint training across partially incompatible datasets. While the marine datasets are forced by atmospheric fields consistent with those used in training, the atmospheric reanalysis ERA5 (used for pre-training) and the operational atmospheric analysis (used for fine-tuning) are forced by different ocean representations that do not match the marine training datasets. In particular, ERA5 is not forced by ORAS6 but by the Operational Sea Surface Temperature and Ice Analysis (OSTIA; \citet{Good2020}). As of today (early 2026), the operational analysis is partially forced by OSTIA, with a delay of up to 1 day, and partially forced by the OCEAN5 initial conditions developed at ECMWF. 
This mismatch in ocean information likely introduces inconsistencies that affect near-surface atmospheric variables. These inconsistencies may contribute to degraded forecast errors in lower-tropospheric pressure-level fields (Fig.~\ref{fig:scorecard_ocean}).
In contrast, we do not observe a comparable degradation in upper-level fields, where ocean influence is weaker at medium-range lead times.
Alongside the degradation for near-surface atmospheric fields, we observe that models with an explicit surface ocean representation better capture the spectral distribution  (Fig.~\ref{fig:spectra}). While this is generally desirable—particularly for RMSE-based training, which tends to smooth spatial structures—the precise origin and implications of this signal require further investigation.

\begin{figure}
    \centering
    \adjustbox{center}{%
        \includegraphics[width=1.0\linewidth]{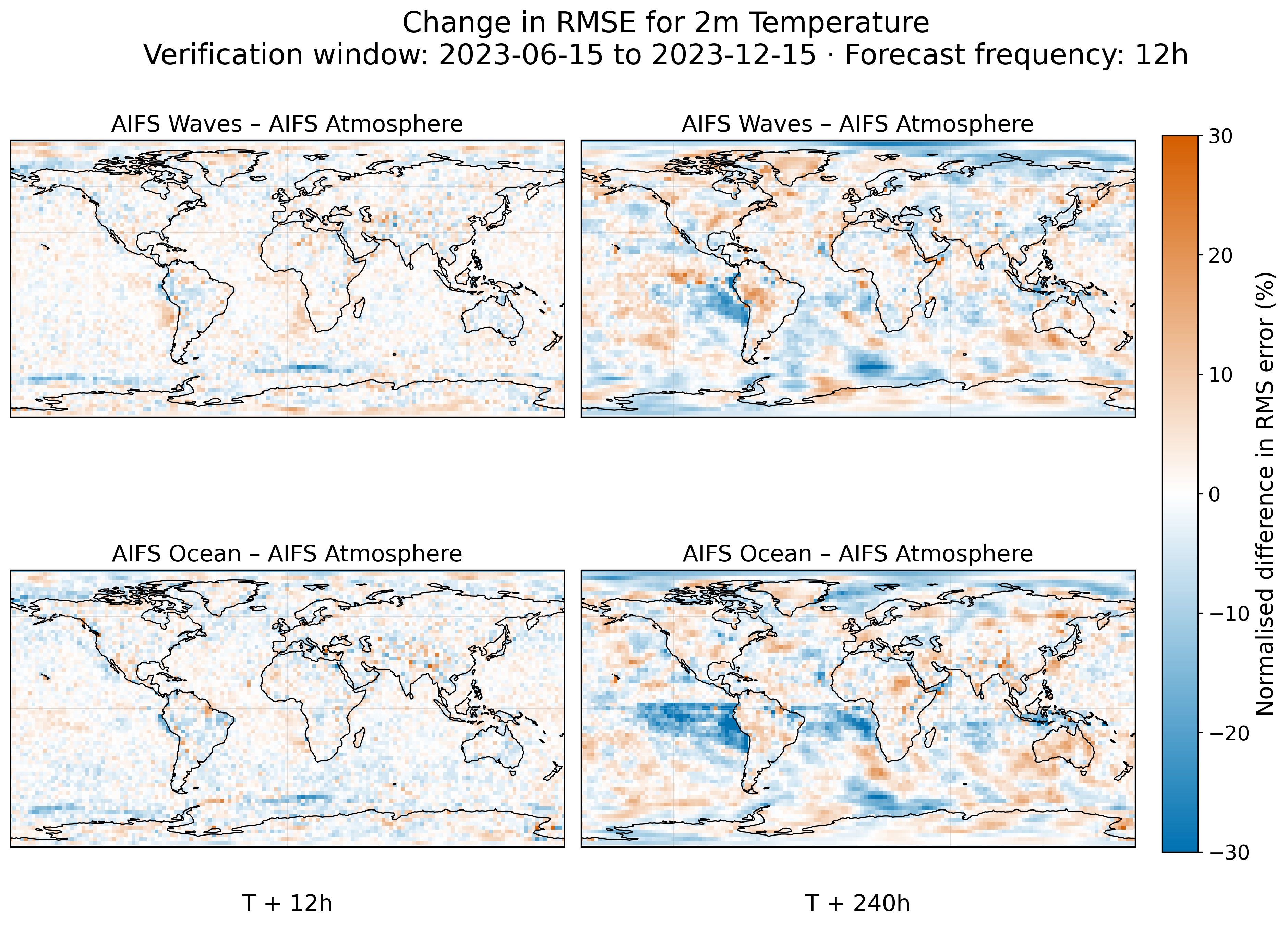}
    }
    \caption{
    Normalised change in RMSE for 2\,m temperature relative to the AIFS Atmosphere for the period 15 June--15 December 2023. Top row shows the impact of the explicit wave representations (with Waves minus no Waves), and bottom row the impact of the explicit ocean representation (with Ocean minus no Ocean). Left panels correspond to lead time T+12\,h and right panels to T+240\,h. Blue colours indicate that adding ocean or wave fields improves the AIFS skill, while red colours indicate a degradation in skill.
    }
    \label{fig:seaicea_2t}
\end{figure}

Sea ice exerts a strong control on near-surface thermodynamics, and this is clearly reflected in the forecast performance at the surface. 
We observe systematic improvements in surface temperature predictions when sea ice is explicitly represented in the model (see Fig.~\ref{fig:seaicea_2t}).  
A weaker but still positive signal is also evident when only waves are included. However, explicit sea ice representation amplifies this benefit. 

In addition, the inclusion of explicit surface ocean representation leads to improvements in tropical temperature skill (see Fig.~\ref{fig:seaicea_2t}). The Tropics are where ocean-atmosphere interaction matters most \cite{PHILANDER1999}, and explicitly evolving sea surface temperature likely provides a more accurate and dynamically responsive lower boundary condition than the implicit oceanic information contained in near-surface atmospheric fields alone. 
In particular, the model can maintain sharper and more coherent SST gradients. These gradients are known to influence near-surface tropical dynamics through their effects on surface fluxes and boundary-layer stability \citep{Lau1997}. An alternative explanation might be that the improved tropical temperature skill is related to a better representation of the diurnal variability of sea surface temperature, which is known to influence tropical convection. More generally, these improvements may reflect a more consistent representation of air--sea coupling and associated feedbacks, although further analysis would be required to disentangle the relative contributions of these mechanisms. 

While the inclusion of surface ocean and sea ice components has a broadly neutral effect on surface temperature scores when comparing against synoptic observations (Fig.~\ref{fig:sfc_obs}, left), we do observe a degradation in surface wind skill.
This suggests that near-surface momentum-related fields may be particularly sensitive to cross-component dataset inconsistencies and competition within the joint training framework (Fig.~\ref{fig:sfc_obs}, right). 
This highlights the importance of dataset coherence and balanced loss design when extending joint models to strongly coupled near-surface processes.

\subsection{Coupled Case Studies}
\label{sec:coupled_test_cases}

Beyond aggregate skill metrics, coupled case studies allow us to examine whether the joint model produces physically coherent cross-component responses in dynamically active and highly coupled regimes.

\begin{figure}
    \centering
    \adjustbox{center}{%
        \includegraphics[width=0.95\linewidth]{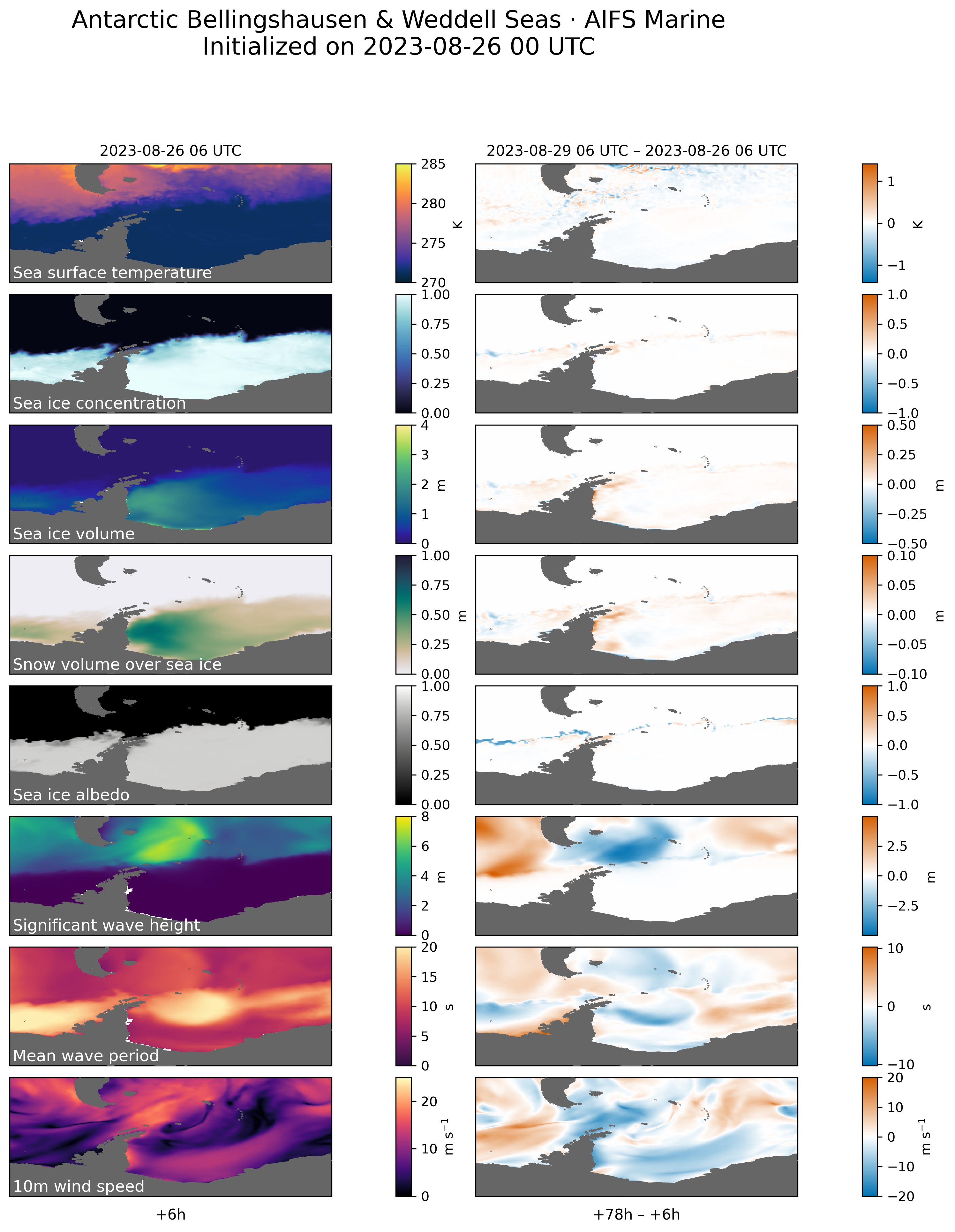}
    }
    \caption{
    AIFS Marine forecast fields for the Bellingshausen and Weddell Seas, initialised on 26 August 2023 at 00 UTC. The left column shows forecasts valid at +6\,h, while the right column shows the difference between the forecasts at +78\,h and +6\,h. Rows show (from top to bottom) sea surface temperature, sea ice concentration, sea ice volume, snow volume over sea ice, sea ice albedo, significant wave height, mean wave period, and 10\,m wind speed.
    }
    \label{fig:sea_ice_waves}
\end{figure}

In the AIFS Marine model, waves, surface ocean, and sea ice fields can be analysed jointly (see Fig.~\ref{fig:sea_ice_waves}). 
Wind waves and wave swell are clearly attenuated in regions covered by sea ice, with wave energy decreasing sharply across the ice edge. 
This behaviour is physically consistent and reflects the damping effect of sea ice on wave propagation, which the model learns implicitly from the training data. 
In addition, the model exhibits coherent interactions between sea ice and the surface ocean: sea ice formation is confined to regions of sufficiently cold sea surface temperatures, and the evolution of sea ice concentration, volume, and related variables remains dynamically consistent with the underlying ocean state. 
Together, these features indicate that the joint model captures key cross-component relationships governing wave attenuation and sea ice thermodynamics without the need for explicitly prescribed coupling mechanisms.

\begin{figure}
    \centering
    \adjustbox{center}{%
        \includegraphics[width=1.0\linewidth]{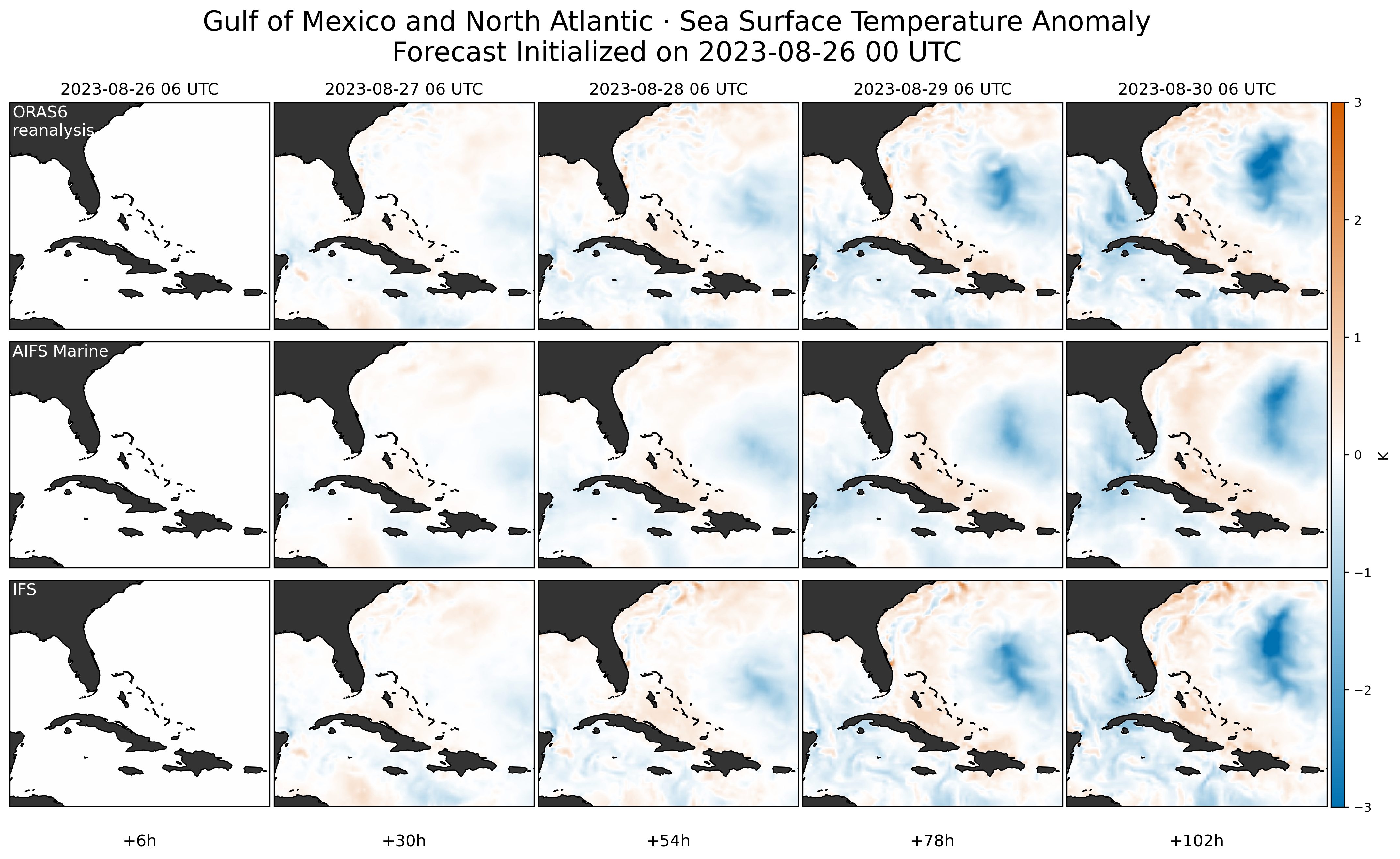}
    }
    \caption{
    Sea surface temperature (SST) anomaly over the Gulf of Mexico and the western North Atlantic for forecasts initialised on 26 August 2023 at 00 UTC. Rows show ORAS6 reanalysis (top), AIFS Marine (middle), and IFS (bottom). Columns correspond to forecast lead times from +6 to +102\,h. SST anomalies are computed relative to the +6\,h field of the respective product. Negative anomalies highlight the cold wakes induced by Hurricanes Idalia (left) and Franklin (centre).
    }
    \label{fig:tc_sst_anomaly}
\end{figure}

To further assess the physical consistency of the joint modelling approach, we analyse a tropical cyclone case study using forecasts from the AIFS Marine model and the IFS. 
We evaluate whether the predicted fields across different Earth system components evolve in a physically consistent way over time.
A visual inspection of Fig.~\ref{fig:tc_AIFS_IFS} reveals a high degree of alignment between atmospheric and marine responses. Regions of strong near-surface winds coincide with enhanced wave activity. In addition, the tropical cyclone imprint is visible in the sea surface temperature, where the passage of the storms induces negative sea surface temperature anomalies (or cold wakes; Fig.~\ref{fig:tc_sst_anomaly}) consistent with the expected wind-driven mixing of colder sub-surface water with the warm surface water.

The ocean response to tropical cyclones further manifests in the sea surface height field. The cyclonic circulation pushes water towards the west coast of Florida, producing a positive sea-surface-height anomaly and increased significant wave height along the coastline. Together, these features are consistent with storm surge. Across all examined variables, the spatial patterns and their cross-component relationships are comparable to those produced by the physics-based IFS forecasts in Fig.~\ref{fig:tc_AIFS_IFS}, indicating that the joint model is able to learn and intrinsically capture key coupled processes governing tropical cyclone evolution.

\subsection{Sensitivity Experiments}

For the atmosphere, perturbation experiments have previously been used to assess physically consistent model behaviour \citep{Hakim2024}. 
Here, we extend this approach to the marine components to complement the quantitative skill assessment. 
We perform a set of sensitivity experiments to evaluate the physical consistency and robustness of the joint modelling framework under strong, out-of-distribution perturbations. 
These experiments probe whether the coupled system responds in a physically meaningful way when key components of the initial state are deliberately modified or removed. 
In contrast to \citet{Hakim2024}, we consider more extreme perturbations, including initial conditions that lie outside the training distribution and extend into physically unrealistic regimes.

\subsubsection{Wave Response to Idealised Initial Conditions}

As a first sensitivity experiment, we examine the response of AIFS Waves to a highly idealised perturbation of the initial state, in line with idealised test cases used in physics-based wave modelling \citep{Hell2025}.
To this end, spatially localised swell energy perturbations are introduced in regions without active weather systems while the rest of the ocean is set to calm conditions.
For this setup, a short ecWAM hindcast is run without wind forcing.
The resulting perturbed wave fields are then used to initialise AIFS Waves.
This setup is not intended to represent a realistic forecast scenario. 
Instead, it serves as a controlled stress test of the model’s ability to propagate wave energy and to generate new wave systems through wind forcing with a response that is well understood analytically.

\begin{figure}
    \centering
    \adjustbox{center}{%
        \includegraphics[width=1.0\linewidth]{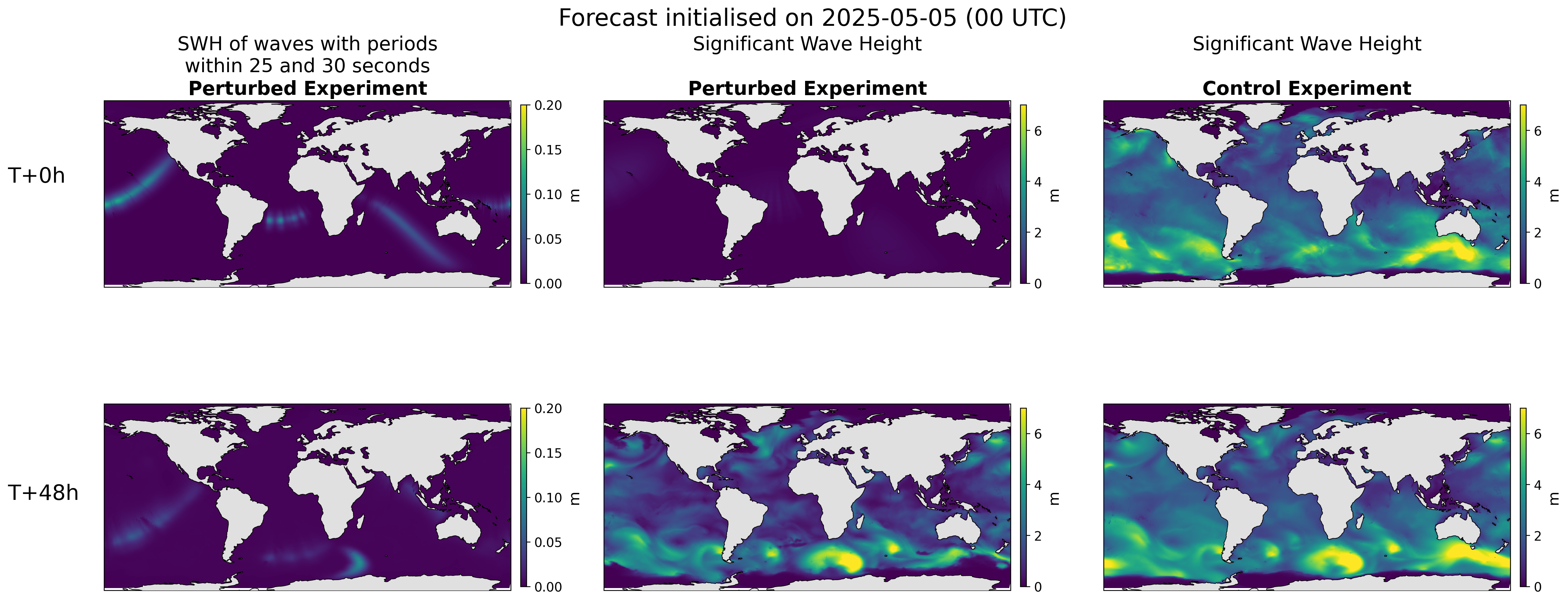}
    }
    \caption{
Initialisation of the AIFS Waves model with synthetic large-period waves at isolated locations, while the remainder of the ocean is set to calm conditions.
Large-period waves with periods within 25 - 30 s propagate over the ocean basins (left), while new wave systems are generated through wind forcing (middle).
These are consistent with the wave patterns obtained from a forecast initialised with unperturbed initial conditions (right).
Nevertheless, without the implicit sea ice mask in the initial state, the sea ice edge is not well captured in the perturbed experiment.
The figure shows the initial states (top) and the forecast after 2 days (bottom).
    }
    \label{fig:wave_stress_test}
\end{figure}

Figure~\ref{fig:wave_stress_test} shows the subsequent evolution of the wave field. 
Large-period swells propagate coherently across ocean basins, as expected from wave dynamics, while new wave systems are generated through wind forcing. 
The resulting wave patterns are consistent with those obtained from forecasts initialised with the unperturbed initial conditions.
This indicates that realistic wave behaviour emerges from the model dynamics rather than being inherited from the initial state.

Nevertheless, the implicit representation of the sea ice edge is degraded in the perturbed forecasts. 
The AIFS Waves model does not include an explicit sea ice component, as it is trained only on atmospheric and wave variables, and therefore lacks a direct mechanism to represent wave attenuation by sea ice. 
In particular, the Antarctic sea ice edge visible in the significant wave height field is displaced too far northward.
This highlights the importance of consistent sea ice information for accurately capturing wave behaviour in polar regions.

Together with the spatial analyses in Sec.~\ref{sec:coupled_test_cases}, this experiment shows that the joint atmosphere–wave model AIFS Waves captures key physical processes. 
These include wave propagation, wind–wave generation, and consistency between surface wind and wave fields, even when the model is initialised far outside the training distribution. 
In this specific case, the training dataset does not include conditions where swell is present in the absence of wind-driven waves.

\subsubsection{Removing Sea Ice from Initial Conditions}

As a complementary sensitivity experiment, we assess the response of the joint AIFS Marine model to the complete removal of sea ice from the initial conditions. 
All sea-ice fields are set to zero in the initial conditions at time $t_0 - 6\mathrm{h}$ and $t_0$, and a small adjustment is applied to the sea-surface temperature (SST) to reduce residual sea-ice memory. 
This adjustment takes the form $+2\,\mathrm{K}\cdot \mathrm{siconc}$, where $\mathrm{siconc}$ denotes the original sea ice concentration prior to removal, thereby preferentially warming regions that were initially ice-covered. 
We compare the subsequent two-month evolution of the perturbed forecasts with control simulations. Differences in sea-ice extent and volume are interpreted as measures of the model’s dynamical and thermodynamic adjustment to the absence of initial sea ice, rather than as conventional forecast verification.

\begin{figure}
    \centering
    \adjustbox{center}{%
        \includegraphics[width=1.1\linewidth]{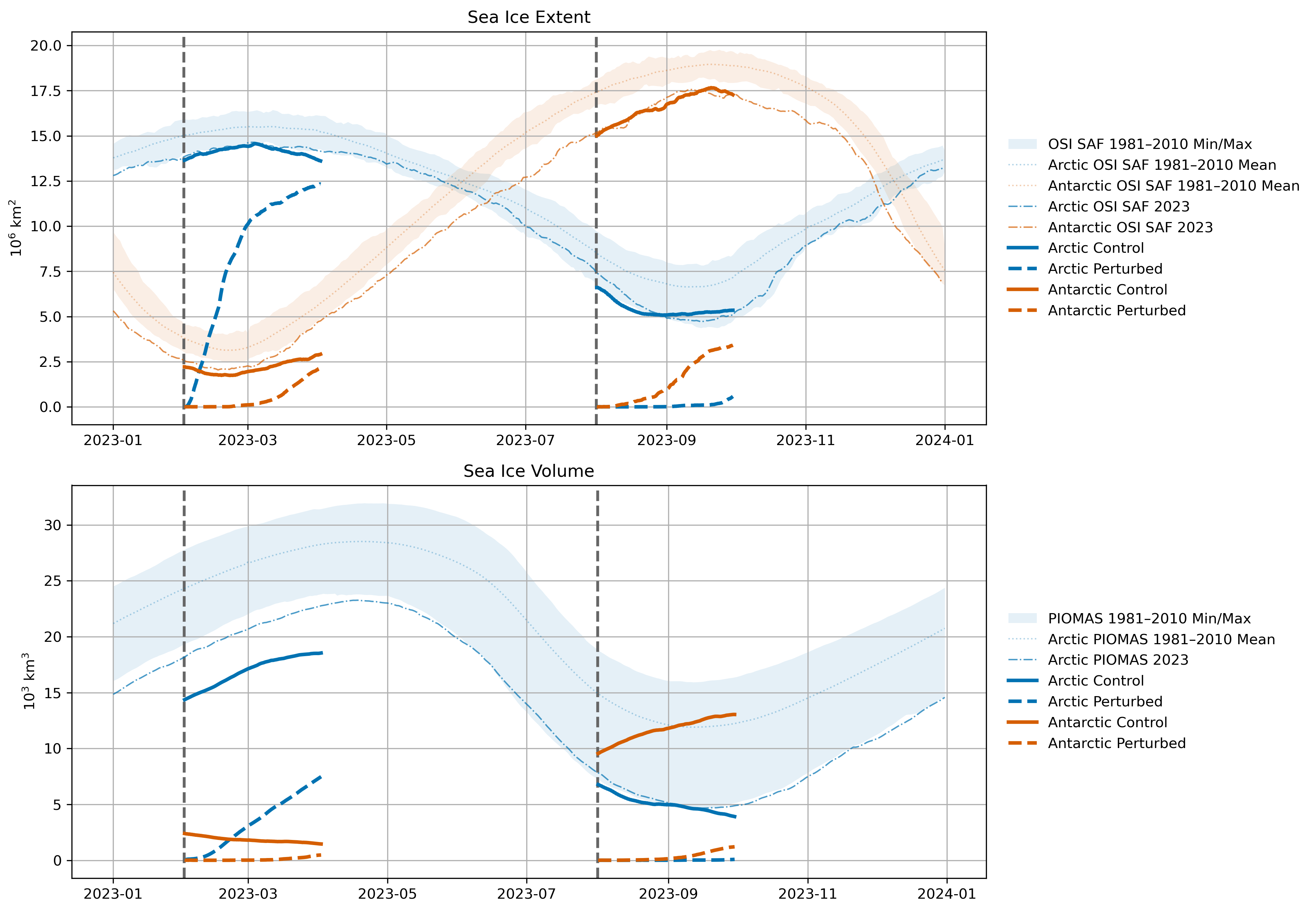}
    }
    \caption{
    Sensitivity experiment in which sea ice is removed from the initial conditions. 
    Time series of (top) sea ice extent and (bottom) sea ice volume for the Arctic (blue) and Antarctic (orange). 
    Solid lines denote control forecasts, while dashed lines show perturbed forecasts initialised without sea ice. 
    Vertical dashed lines indicate the two forecast initialisation dates (1 February 2023 and 1 August 2023). 
    Shaded envelopes and dotted lines show the OSI SAF (extent) and PIOMAS (volume, Arctic only) 1981--2010 climatological range and mean, respectively, with dash-dotted lines indicating the corresponding 2023 observational estimates.
    }
    \label{fig:seaice_sensitivity_timeseries}
\end{figure}

Fig.~\ref{fig:seaice_sensitivity_timeseries} illustrates the hemispheric response of the coupled system to the removal of sea ice from the initial conditions, shown in terms of sea ice extent and volume for both the Arctic and Antarctic, and compared to the extent and volume indices from OSI-SAF (Ocean and Sea Ice Satellite Application Facility; \citet{OSI-SAF}) and PIOMAS (Pan-Arctic Ice Ocean Modeling and Assimilation System; \citet{Schweiger2011}), respectively. 
In all cases, sea ice does not reappear instantaneously in the perturbed forecasts, but instead recovers gradually, indicating that the model does not trivially reconstruct ice from residual memory in other state variables. 
For the February initialisation, ice regrowth first occurs in the Arctic, where thermodynamic conditions favour wintertime freezing.
The Antarctic, on the other hand, remains largely ice-free until later in the forecast, as the austral seasonal cycle reaches freeze-up (September). 
Conversely, for the August initialisation, sea ice recovery is initially confined to the Antarctic, consistent with the onset of austral freeze-up, while Arctic ice formation is delayed until colder conditions develop later in the forecast period. 
This seasonally asymmetric behaviour demonstrates that the joint model responds primarily to the evolving large-scale thermodynamic state rather than to the imposed perturbation itself.

Across both hemispheres and initialisation dates, the recovery of sea ice volume is systematically slower than that of sea ice extent. 
While extent increases relatively rapidly once freezing conditions are re-established, volume builds up more gradually, reflecting the longer timescales associated with ice thickening. 
The separation of timescales between extent and volume is physically expected and consistent with the spin-up behaviour seen in numerical coupled models \citep{Schrder2007, Tietsche2011}. 
This provides further evidence that the joint AIFS system adjusts realistically, both dynamically and thermodynamically, to the absence of initial sea ice.
Overall, the recovery of sea ice extent could in principle occur more quickly; however, in the perturbed forecasts, this recovery is delayed by the imposed $+2\,\mathrm{K}$ sea-surface temperature adjustment, which requires the system first to transition back to freezing conditions. 

Fig.~\ref{fig:seaice_sensitivity_onset} provides a more detailed view of the Arctic sea ice recovery in the perturbed forecast initialised on 1 February 2023, focusing on the spatial structure and temporal characteristics of ice regrowth. 
The top panel shows the evolution of grid-point-averaged sea ice volume per unit area, aligned to the local onset of ice formation, defined by a combined threshold in sea ice volume and concentration. 
Following the onset, sea ice volume increases rapidly during the first days.
Then it exhibits a progressively slower growth rate over the subsequent weeks, with substantial spatial variability reflected by the spread across grid points.
The bottom panels illustrate the spatial evolution of the ice state, with the left panel confirming the imposed ice-free initial condition, the centre panel showing the recovered sea ice volume after 30 days, and the right panel indicating the geographical distribution of freeze-up dates across the Arctic basin.
Together, these diagnostics characterise both the timing and spatial heterogeneity of sea ice regrowth following the removal of the initial sea ice state.

The temporal and spatial characteristics of the recovered sea ice are physically consistent with known thermodynamic behaviour. 
The rapid initial increase in sea ice volume following onset, followed by a gradual slowdown, reflects the self-limiting nature of thermodynamic ice growth: as ice thickens and snow accumulates on its surface, the insulating effect reduces conductive heat loss from the ocean, inhibiting further growth \citep{Zampieri2024}. 
The magnitude of the recovered ice volume remains within the range expected for winter sea ice conditions, indicating that the model does not produce unrealistically rapid or excessive regrowth despite the strong initial perturbation.
Spatially, ice formation occurs earlier and more robustly in peripheral Arctic seas, particularly in regions adjacent to cold continental landmasses and over shallow bathymetry, where the upper ocean can cool more efficiently. 
These regions correspond to well-known hotspots of observed sea ice formation, lending further credibility to the simulated recovery patterns.
Taken together, the timing, growth rates, and spatial distribution of the regrown ice suggest that the joint AIFS Marine system responds to the perturbation in a physically meaningful manner, with sea ice evolution emerging from large-scale thermodynamic forcing rather than from residual memory of the removed initial state.

\section{Discussion and Conclusion}

This study demonstrates that joint ML models can successfully provide a more complete representation of the Earth system by simultaneously forecasting atmospheric and marine components even if they evolve on distinct spatial and temporal scales. 
Joint modelling challenges the paradigm of traditional physics-based forecasting systems, where the different Earth system subcomponents have to be simulated by distinct models which are then explicitly coupled, a procedure that intrinsically generates numerical errors \citep{Gross2018, Schller2025}.
For example, in the ECMWF forecasting system, the atmospheric component (IFS) runs at a horizontal resolution of about 9 km with a time step of 450 s, while the ocean component (NEMO4) operates on a coarser grid of about 25 km with a time step of 1200 s, with the two components exchanging fields at hourly coupling intervals.
A key result of the joint approach is that coupling between Earth system components does not need to be prescribed explicitly in machine-learned models, as similarly observed in \citep{Boucher2025}.
Instead, information exchange is modelled intrinsically and emerges naturally through the shared latent representation, allowing the model to learn cross-component dependencies directly from data.

The joint models produce skilful forecasts for the surface ocean, sea ice, and ocean wave components. 
For nearly all evaluated marine variables—sea surface height being the notable exception—we observe an improvement of approximately one day in forecast skill at medium-range lead times compared to physics-based models. 
In practical terms, this corresponds to users gaining actionable information about marine conditions roughly one day earlier. 
The magnitude of this improvement is comparable to that previously reported for atmospheric forecasts in AIFS \citep{Lang2024a}.
In addition, ML forecasts are available earlier than those from the physics-based model due to the faster inference of the ML system.

A central question addressed by this work concerns the relative roles of implicit and explicit representation in ML Earth system models.
The atmospheric-only AIFS achieves competitive forecast skill without explicit marine inputs, suggesting that oceanic and cryospheric effects implicitly contained in reanalysis datasets are sufficient at medium-range lead times.
This contrasts with physics-based modelling, where the absence of an interactive ocean is known to degrade forecast quality at similar lead times \citep{Graham2005, Brassington2015, Berthou2016, Vellinga2020, Gurmy2005, Berthou2025}.

However, our results also show that explicit representation becomes increasingly beneficial at longer lead times.
Surface temperature skill improves in the marginal ice zone when wave fields are included.
Larger gains (15\% RMSE reduction) are obtained when sea ice is represented explicitly, with similar improvements observed for wave dynamics near the ice edge.
This points to an increasing importance of explicit ocean representation at subseasonal and seasonal timescales, where memory effects become more influential and a deep ocean representation may be required.

From a numerical perspective, the joint models are stable over medium-range timescales and beyond, as demonstrated by the sea ice case study, with no evidence of artefacts or instabilities in the newly introduced marine fields. 
The models also remain robust under strong perturbations of the initial conditions and for inputs outside the training distribution, supporting their suitability for operational use.

Physical consistency in the joint models is achieved through a combination of explicit constraints and intrinsically learned relationships.
While bounding and post-processing are used to enforce physically admissible ranges, for some variables, we favour leaky constraints during training.
These ensure numerical stability without inducing vanishing gradients in bounded regions of the state space. 
The models also learn physically meaningful relationships directly from the data, as evidenced by coherent cross-component behaviour in extreme events.
Evaluating model performance under rare and out-of-distribution scenarios—such as configurations with absent sea ice—provides a valuable stress test and is essential for building trust in ML-based weather forecasting systems, particularly as the climate system evolves beyond historical conditions.
At the same time, the systematic underestimation of sea surface height points to limitations linked to the mean climate state represented during the training period, indicating that further investigation is required.
These technical developments have been implemented within the Anemoi software ecosystem, which provides the infrastructure for dataset preparation, model training, and inference across Earth system components.

Dataset consistency is a key requirement for the joint modelling approach across Earth system components. 
Small mismatches between atmospheric and oceanic datasets likely contribute to the small degradations observed in some atmospheric fields when marine components are added.
One way to address this is through improved dataset alignment, possibly by training on coupled reanalyses, albeit at substantial computational and development cost.

A complementary and promising approach under parallel development \citep{DestinE2025Waves, DestinE2026SeaIce} is the coupling of component-wise models, inspired by established strategies in physics-based systems. 
Coupling enables different components to operate on their native temporal resolutions while exchanging information dynamically, and is expected to reduce negative impacts on atmospheric forecast quality.
As with joint modelling, coupling requires temporally consistent data coverage across components. 
In this work, for example, this meant reducing the training period, since no ocean reanalysis was available before 1993.
More flexible foundational modelling approaches, such as that proposed by \citet{Bodnar2025}, provide an additional pathway by relaxing some of the structural constraints imposed by the requirement for full data coverage, while still allowing the model to learn cross-component couplings directly from the data.

Next steps will include extending the joint modelling approach to probabilistic ML models to explicitly represent forecast uncertainty, for example, through proper-score optimisation \citep{Lang2024b} or diffusion-based approaches \citep{Price2023}. 
Initial results from probabilistic joint models incorporating surface-ocean and sea-ice components are already being evaluated, including the sub-seasonal AIFS-Thalassa model \citep{AIFSThalassa} for the AI Weather Quest \citep{Loegel2025}, indicating that joint probabilistic Earth system modelling is both feasible and promising.

\bibliography{sample}

@article{Hersbach2020,
  title = {The {ERA5} global reanalysis},
  volume = {146},
  ISSN = {1477-870X},
  url = {http://dx.doi.org/10.1002/qj.3803},
  DOI = {10.1002/qj.3803},
  number = {730},
  journal = {Quarterly Journal of the Royal Meteorological Society},
  publisher = {Wiley},
  author = {Hersbach,  Hans and Bell,  Bill and Berrisford,  Paul and Hirahara,  Shoji and Horányi,  András and Muñoz‐Sabater,  Joaquín and Nicolas,  Julien and Peubey,  Carole and Radu,  Raluca and Schepers,  Dinand and Simmons,  Adrian and Soci,  Cornel and Abdalla,  Saleh and Abellan,  Xavier and Balsamo,  Gianpaolo and Bechtold,  Peter and Biavati,  Gionata and Bidlot,  Jean and Bonavita,  Massimo and De Chiara,  Giovanna and Dahlgren,  Per and Dee,  Dick and Diamantakis,  Michail and Dragani,  Rossana and Flemming,  Johannes and Forbes,  Richard and Fuentes,  Manuel and Geer,  Alan and Haimberger,  Leo and Healy,  Sean and Hogan,  Robin J. and Hólm,  Elías and Janisková,  Marta and Keeley,  Sarah and Laloyaux,  Patrick and Lopez,  Philippe and Lupu,  Cristina and Radnoti,  Gabor and de Rosnay,  Patricia and Rozum,  Iryna and Vamborg,  Freja and Villaume,  Sebastien and Thépaut,  Jean‐Noël},
  year = {2020},
  month = jun,
  pages = {1999–2049}
}

@article{Lam2023,
author = {Remi Lam  and Alvaro Sanchez-Gonzalez  and Matthew Willson  and Peter Wirnsberger  and Meire Fortunato  and Ferran Alet  and Suman Ravuri  and Timo Ewalds  and Zach Eaton-Rosen  and Weihua Hu  and Alexander Merose  and Stephan Hoyer  and George Holland  and Oriol Vinyals  and Jacklynn Stott  and Alexander Pritzel  and Shakir Mohamed  and Peter Battaglia },
title = {Learning skillful medium-range global weather forecasting},
journal = {Science},
volume = {382},
number = {6677},
pages = {1416-1421},
year = {2023},
doi = {10.1126/science.adi2336},
URL = {https://www.science.org/doi/abs/10.1126/science.adi2336},
eprint = {https://www.science.org/doi/pdf/10.1126/science.adi2336},}

@misc{Keisler2022,
      title={{Forecasting Global Weather with Graph Neural Networks}}, 
      author={Ryan Keisler},
      year={2022},
      eprint={2202.07575},
      archivePrefix={arXiv},
      primaryClass={physics.ao-ph},
      url={https://arxiv.org/abs/2202.07575}, 
}

@misc{Pathak2022,
      title={{FourCastNet}: A Global Data-driven High-resolution Weather Model using Adaptive Fourier Neural Operators}, 
      author={Jaideep Pathak and Shashank Subramanian and Peter Harrington and Sanjeev Raja and Ashesh Chattopadhyay and Morteza Mardani and Thorsten Kurth and David Hall and Zongyi Li and Kamyar Azizzadenesheli and Pedram Hassanzadeh and Karthik Kashinath and Animashree Anandkumar},
      year={2022},
      eprint={2202.11214},
      archivePrefix={arXiv},
      primaryClass={physics.ao-ph},
      url={https://arxiv.org/abs/2202.11214}, 
}

@article{Bi2023,
title = {Accurate medium-range global weather forecasting with 3D neural networks},
author = {Bi, Kaifeng and Xie, Lei and Zhang, Hao and others},
journal = {Nature},
volume = {619},
pages = {533--538},
year = {2023},
doi = {10.1038/s41586-023-06185-3},
url = {https://doi.org/10.1038/s41586-023-06185-3}
}

@article{Chen2023,
title = {{FuXi}: a cascade machine learning forecasting system for 15-day global weather forecast},
author = {Chen, Lei and Zhong, Xinyu and Zhang, Feng and Cheng, Yuan and Xu, Yinghui and Qi, Yuan and Li, Hao},
journal = {npj Climate and Atmospheric Science},
volume = {6},
pages = {190},
year = {2023},
doi = {10.1038/s41612-023-00512-1},
url = {https://doi.org/10.1038/s41612-023-00512-1}
}

@misc{Lang2024a,
  doi = {10.48550/ARXIV.2406.01465},
  url = {https://arxiv.org/abs/2406.01465},
  author = {Lang,  Simon and Alexe,  Mihai and Chantry,  Matthew and Dramsch,  Jesper and Pinault,  Florian and Raoult,  Baudouin and Clare,  Mariana C. A. and Lessig,  Christian and Maier-Gerber,  Michael and Magnusson,  Linus and Bouallègue,  Zied Ben and Nemesio,  Ana Prieto and Dueben,  Peter D. and Brown,  Andrew and Pappenberger,  Florian and Rabier,  Florence},
  title = {{AIFS} -- {ECMWF}'s data-driven forecasting system},
  publisher = {arXiv},
  year = {2024},
  copyright = {Creative Commons Attribution Share Alike 4.0 International}
}

@article{Aouni2025,
  author = {El Aouni, Anass and Gaudel, Quentin and Regnier, Charly and Van Gennip, Simon and Le Galloudec, Olivier and Drevillon, Marie and Drillet, Yann and Lellouche, Jean-Michel},
  title = {{GLONET: Mercator's End-to-End Neural Global Ocean Forecasting System}},
  journal = {Journal of Geophysical Research: Machine Learning and Computation},
  volume = {2},
  number = {3},
  pages = {e2025JH000686},
  doi = {https://doi.org/10.1029/2025JH000686},
  url = {https://agupubs.onlinelibrary.wiley.com/doi/abs/10.1029/2025JH000686},
  eprint = {https://agupubs.onlinelibrary.wiley.com/doi/pdf/10.1029/2025JH000686},
  note = {e2025JH000686 2025JH000686},
  year = {2025}
}

@misc{Wang2024,
      title={{XiHe: A Data-Driven Model for Global Ocean Eddy-Resolving Forecasting}}, 
      author={Xiang Wang and Renzhi Wang and Ningzi Hu and Pinqiang Wang and Peng Huo and Guihua Wang and Huizan Wang and Senzhang Wang and Junxing Zhu and Jianbo Xu and Jun Yin and Senliang Bao and Ciqiang Luo and Ziqing Zu and Yi Han and Weimin Zhang and Kaijun Ren and Kefeng Deng and Junqiang Song},
      year={2024},
      eprint={2402.02995},
      archivePrefix={arXiv},
      primaryClass={physics.ao-ph},
      url={https://arxiv.org/abs/2402.02995}, 
}

@article{Wedi2014,
  title = {Increasing horizontal resolution in numerical weather prediction and climate simulations: illusion or panacea?},
  volume = {372},
  ISSN = {1471-2962},
  url = {http://dx.doi.org/10.1098/rsta.2013.0289},
  DOI = {10.1098/rsta.2013.0289},
  number = {2018},
  journal = {Philosophical Transactions of the Royal Society A: Mathematical,  Physical and Engineering Sciences},
  publisher = {The Royal Society},
  author = {Wedi,  Nils P.},
  year = {2014},
  month = jun,
  pages = {20130289}
}

@article{Zuo2024,
  doi = {10.21957/HZD5Y821LK},
  url = {https://www.ecmwf.int/en/elibrary/81576-ecmwfs-next-ensemble-reanalysis-system-ocean-and-sea-ice-oras6},
  author = {Zuo,  Hao and Alonso Balmaseda,  Magdalena and de Boisseson,  Eric and Browne,  Philip and Chrust,  Marcin and Keeley,  Sarah and Mogensen,  Kristian and Pelletier,  Charles and de Rosnay,  Patricia and Takakura,  Toshinari},
  title = {{ECMWF}’s next ensemble reanalysis system for ocean and sea ice: {ORAS6}},
  publisher = {{ECMWF}},
  year = {2024}
}

@article{Lipscomb2001,
  title = {Remapping the thickness distribution in sea ice models},
  volume = {106},
  ISSN = {0148-0227},
  url = {http://dx.doi.org/10.1029/2000JC000518},
  DOI = {10.1029/2000jc000518},
  number = {C7},
  journal = {Journal of Geophysical Research: Oceans},
  publisher = {American Geophysical Union (AGU)},
  author = {Lipscomb,  William H.},
  year = {2001},
  month = jul,
  pages = {13989–14000}
}

@article{Massonnet2019,
  title = {On the discretization of the ice thickness distribution  in the  {NEMO3.6-LIM3} global ocean–sea ice model},
  volume = {12},
  ISSN = {1991-9603},
  url = {http://dx.doi.org/10.5194/gmd-12-3745-2019},
  DOI = {10.5194/gmd-12-3745-2019},
  number = {8},
  journal = {Geoscientific Model Development},
  publisher = {Copernicus GmbH},
  author = {Massonnet,  Fran\c{c}ois and Barthélemy,  Antoine and Worou,  Koffi and Fichefet,  Thierry and Vancoppenolle,  Martin and Rousset,  Clément and Moreno-Chamarro,  Eduardo},
  year = {2019},
  month = aug,
  pages = {3745–3758}
}

@article{Vancoppenolle2009,
  title = {Simulating the mass balance and salinity of Arctic and Antarctic sea ice. 1. Model description and validation},
  volume = {27},
  ISSN = {1463-5003},
  url = {http://dx.doi.org/10.1016/j.ocemod.2008.10.005},
  DOI = {10.1016/j.ocemod.2008.10.005},
  number = {1–2},
  journal = {Ocean Modelling},
  publisher = {Elsevier BV},
  author = {Vancoppenolle,  Martin and Fichefet,  Thierry and Goosse,  Hugues and Bouillon,  Sylvain and Madec,  Gurvan and Maqueda,  Miguel Angel Morales},
  year = {2009},
  pages = {33–53}
}

@techreport{ECMWF2024IFS,
  author      = {{ECMWF}},
  title       = {{{IFS} Documentation – Cy49r1: Part VII: {ECMWF} Wave Model}},
  institution = {European Centre for Medium-Range Weather Forecasts},
  address     = {Reading, UK},
  year        = {2024},
  note        = {Operational implementation 12 November 2024},
  url         = {https://www.ecmwf.int/sites/default/files/elibrary/112024/81629-ifs-documentation-cy49r1-part-vii-ecmwf-wave-model.pdf},
}

@article{Yu2022,
title = {A new method for parameterization of wave dissipation by sea ice},
journal = {Cold Regions Science and Technology},
volume = {199},
pages = {103582},
year = {2022},
issn = {0165-232X},
doi = {10.1016/j.coldregions.2022.103582},
url = {https://www.sciencedirect.com/science/article/pii/S0165232X2200101X},
author = {Jie Yu and W. Erick Rogers and David W. Wang},
}

@misc{Boucher2025,
      title={Learning Coupled Earth System Dynamics with GraphDOP}, 
      author={Eulalie Boucher and Mihai Alexe and Peter Lean and Ewan Pinnington and Simon Lang and Patrick Laloyaux and Lorenzo Zampieri and Patricia de Rosnay and Niels Bormann and Anthony McNally},
      year={2025},
      eprint={2510.20416},
      archivePrefix={arXiv},
      primaryClass={physics.ao-ph},
      url={https://arxiv.org/abs/2510.20416}, 
}

@article{Moldovan2025,
  title = {{AIFS} 1.1.0: An update to {ECMWF}’s machine-learned weather forecast model {AIFS}},
  url = {http://dx.doi.org/10.5194/egusphere-2025-4716},
  DOI = {10.5194/egusphere-2025-4716},
  publisher = {Copernicus GmbH},
  author = {Moldovan,  Gabriel and Pinnington,  Ewan and Prieto Nemesio,  Ana and Lang,  Simon and Ben Bouallègue,  Zied and Dramsch,  Jesper and Alexe,  Mihai and Santa Cruz,  Mario and Hahner,  Sara and Cook,  Harrison and Theissen,  Helen and Clare,  Mariana and O’Brien,  Cathal and Polster,  Jan and Magnusson,  Linus and Mertes,  Gert and Pinault,  Florian and Raoult,  Baudouin and de Rosnay,  Patricia and Forbes,  Richard and Chantry,  Matthew},
  year = {2025},
  month = oct 
}

@misc{Lang2024b,
      title={{AIFS-CRPS}: Ensemble forecasting using a model trained with a loss function based on the Continuous Ranked Probability Score}, 
      author={Simon Lang and Mihai Alexe and Mariana C. A. Clare and Christopher Roberts and Rilwan Adewoyin and Zied Ben Bouallègue and Matthew Chantry and Jesper Dramsch and Peter D. Dueben and Sara Hahner and Pedro Maciel and Ana Prieto-Nemesio and Cathal O'Brien and Florian Pinault and Jan Polster and Baudouin Raoult and Steffen Tietsche and Martin Leutbecher},
      year={2024},
      eprint={2412.15832},
      archivePrefix={arXiv},
      primaryClass={physics.ao-ph},
      url={https://arxiv.org/abs/2412.15832}, 
}

@misc{Madec2024,
  author       = {Gurvan Madec and
                  Mike Bell and
                  Rachid Benshila and
                  Adam Blaker and
                  Romain Boudrallé-Badie and
                  Clément Bricaud and
                  Diego Bruciaferri and
                  Davi Carneiro and
                  Miguel Castrillo and
                  Daley Calvert and
                  Jérômeme Chanut and
                  Emanuela Clementi and
                  Andrew Coward and
                  Casimir de Lavergne and
                  Srdan Dobricic and
                  Italo Epicoco and
                  Christian Éthé and
                  Emma Fiedler and
                  David Ford and
                  Rachel Furner and
                  Jonas Ganderton and
                  Tim Graham and
                  James Harle and
                  Katherine Hutchinson and
                  Doroteaciro Iovino and
                  Robert King and
                  Dan Lea and
                  Claire Levy and
                  Tomas Lovato and
                  Eric Maisonnave and
                  Julian Mak and
                  Juan Manuel Castillo Sanchez and
                  Matt Martin and
                  Nicolas Martin and
                  Diana Martins and
                  Sébastien Masson and
                  Pierre Mathiot and
                  Francesca Mele and
                  Silvia Mocavero and
                  Aimie Moulin and
                  Simon Müller and
                  George Nurser and
                  Paolo Oddo and
                  Stella Paronuzzi and
                  Julien Paul and
                  Mathieu Peltier and
                  Renaud Person and
                  Clement Rousset and
                  Stefanie Rynders and
                  Guillaume Samson and
                  David Schroeder and
                  Dave Storkey and
                  Andrea Storto and
                  Sibylle Téchené and
                  Martin Vancoppenolle and
                  Chris Wilson},
  title        = {{NEMO} Ocean Engine Reference Manual},
  month        = dec,
  year         = 2024,
  publisher    = {Zenodo},
  version      = {v5.0},
  doi          = {10.5281/zenodo.14515373},
  url          = {https://doi.org/10.5281/zenodo.14515373},
}

@article{Vancoppenolle2023,
  doi = {10.5281/ZENODO.7534900},
  url = {https://zenodo.org/record/7534900},
  author = {Vancoppenolle, Martin and Rousset, Clement and Blockley, Edward and Aksenov, Yevgeny and Feltham, Daniel and Fichefet, Thierry and Garric, Gilles and Guémas, Virginie and Iovino, Dorotea and Keeley, Sarah and Madec, Gurvan and Massonnet, François and Ridley, Jeff and Schroeder, David and Tietsche, Steffen},
  language = {en},
  title = {{SI3},  the {NEMO} Sea Ice Engine},
  publisher = {Zenodo},
  year = {2023},
  copyright = {Creative Commons Attribution 4.0 International}
}

@article{Mogensen2012,
  doi = {10.21957/X5Y9YRTM},
  url = {https://www.ecmwf.int/node/11174},
  author = {Mogensen,  Kristian and Weaver, Anthony and Alonso Balmaseda,  Magdalena},
  title = {The {NEMOVAR} ocean data assimilation system as implemented in the {ECMWF} ocean analysis for System 4},
  publisher = {{ECMWF}},
  year = {2012}
}

@article{PHILANDER1999,
  title = {A review of tropical ocean-atmosphere interactions},
  volume = {51},
  ISSN = {1600-0889},
  url = {http://dx.doi.org/10.1034/j.1600-0889.1999.00007.x},
  DOI = {10.1034/j.1600-0889.1999.00007.x},
  number = {1},
  journal = {Tellus B},
  publisher = {Stockholm University Press},
  author = {Philander,  S. George},
  year = {1999},
  month = Feb,
  pages = {71–90}
}

@article{Bodnar2025,
  title = {A foundation model for the Earth system},
  volume = {641},
  ISSN = {1476-4687},
  url = {http://dx.doi.org/10.1038/s41586-025-09005-y},
  DOI = {10.1038/s41586-025-09005-y},
  number = {8065},
  journal = {Nature},
  publisher = {Springer Science and Business Media LLC},
  author = {Bodnar,  Cristian and Bruinsma,  Wessel P. and Lucic,  Ana and Stanley,  Megan and Allen,  Anna and Brandstetter,  Johannes and Garvan,  Patrick and Riechert,  Maik and Weyn,  Jonathan A. and Dong,  Haiyu and Gupta,  Jayesh K. and Thambiratnam,  Kit and Archibald,  Alexander T. and Wu,  Chun-Chieh and Heider,  Elizabeth and Welling,  Max and Turner,  Richard E. and Perdikaris,  Paris},
  year = {2025},
  month = may,
  pages = {1180–1187}
}

@article{Goessling2016,
  title = {Predictability of the Arctic sea ice edge},
  volume = {43},
  ISSN = {1944-8007},
  url = {http://dx.doi.org/10.1002/2015GL067232},
  DOI = {10.1002/2015gl067232},
  number = {4},
  journal = {Geophysical Research Letters},
  publisher = {American Geophysical Union (AGU)},
author = {Goessling, Helge F. and Tietsche, Steffen and Day, Jonathan J. and Hawkins, Ed and Jung, Thomas},
  year = {2016},
  month = feb,
  pages = {1642–1650}
}

@article{Zampieri2018,
  title = {Bright Prospects for Arctic Sea Ice Prediction on Subseasonal Time Scales},
  volume = {45},
  ISSN = {1944-8007},
  url = {http://dx.doi.org/10.1029/2018GL079394},
  DOI = {10.1029/2018gl079394},
  number = {18},
  journal = {Geophysical Research Letters},
  publisher = {American Geophysical Union (AGU)},
  author = {Zampieri,  Lorenzo and Goessling,  Helge F. and Jung,  Thomas},
  year = {2018},
  month = sep,
  pages = {9731–9738}
}

@article{Zampieri2024,
  title = {Modeling the Winter Heat Conduction Through the Sea Ice System During {MOSAiC}},
  volume = {51},
  ISSN = {1944-8007},
  url = {http://dx.doi.org/10.1029/2023GL106760},
  DOI = {10.1029/2023gl106760},
  number = {8},
  journal = {Geophysical Research Letters},
  publisher = {American Geophysical Union (AGU)},
  author = {Zampieri,  Lorenzo and Clemens‐Sewall,  David and Sledd,  Anne and Hutter,  Nils and Holland,  Marika},
  year = {2024},
  month = apr 
}

@article{Schweiger2011,
  title = {Uncertainty in modeled Arctic sea ice volume},
  volume = {116},
  ISSN = {0148-0227},
  url = {http://dx.doi.org/10.1029/2011JC007084},
  DOI = {10.1029/2011jc007084},
  journal = {Journal of Geophysical Research},
  publisher = {American Geophysical Union (AGU)},
  author = {Schweiger,  Axel and Lindsay,  Ron and Zhang,  Jinlun and Steele,  Mike and Stern,  Harry and Kwok,  Ron},
  year = {2011},
  month = sep 
}

@article{Good2020,
  title = {The Current Configuration of the {OSTIA} System for Operational Production of Foundation Sea Surface Temperature and Ice Concentration Analyses},
  volume = {12},
  ISSN = {2072-4292},
  url = {http://dx.doi.org/10.3390/rs12040720},
  DOI = {10.3390/rs12040720},
  number = {4},
  journal = {Remote Sensing},
  publisher = {MDPI AG},
  author = {Good,  Simon and Fiedler,  Emma and Mao,  Chongyuan and Martin,  Matthew J. and Maycock,  Adam and Reid,  Rebecca and Roberts-Jones,  Jonah and Searle,  Toby and Waters,  Jennifer and While,  James and Worsfold,  Mark},
  year = {2020},
  month = feb,
  pages = {720}
}

@article{Lau1997,
  title = {The Role of Large-Scale Atmospheric Circulation in the Relationship between Tropical Convection and Sea Surface Temperature},
  volume = {10},
  ISSN = {1520-0442},
  url = {http://dx.doi.org/10.1175/1520-0442(1997)010<0381:TROLSA>2.0.CO;2},
  DOI = {10.1175/1520-0442(1997)010<0381:trolsa>2.0.co;2},
  number = {3},
  journal = {Journal of Climate},
  publisher = {American Meteorological Society},
  author = {Lau, William K-M. and Wu,  H-T. and Bony,  Sandrine},
  year = {1997},
  month = mar,
  pages = {381–392}
}

@misc{OSI-SAF,
  doi = {10.15770/EUM_SAF_OSI_0022},
  url = {https://user.eumetsat.int/catalogue/EO:EUM:DAT:0875},
  author = {{OSI SAF} and {EUMETSAT SAF on Ocean and Sea Ice}},
  language = {test},
  title = {Sea Ice Index - Multimission},
  publisher = {EUMETSAT SAF on Ocean and Sea Ice},
  year = {2020},
  copyright = {Free and unrestricted}
}

@article{Bushuk2024,
  title = {Predicting September Arctic Sea Ice: A Multimodel Seasonal Skill Comparison},
  volume = {105},
  ISSN = {1520-0477},
  url = {http://dx.doi.org/10.1175/BAMS-D-23-0163.1},
  DOI = {10.1175/bams-d-23-0163.1},
  number = {7},
  journal = {Bulletin of the American Meteorological Society},
  publisher = {American Meteorological Society},
  author = {Bushuk,  Mitchell and Ali,  Sahara and Bailey,  David A. and Bao,  Qing and Batté,  Lauriane and Bhatt,  Uma S. and Blanchard-Wrigglesworth,  Edward and Blockley,  Ed and Cawley,  Gavin and Chi,  Junhwa and Counillon,  Fran\c{c}ois and Coulombe,  Philippe Goulet and Cullather,  Richard I. and Diebold,  Francis X. and Dirkson,  Arlan and Exarchou,  Eleftheria and G\"{o}bel,  Maximilian and Gregory,  William and Guemas,  Virginie and Hamilton,  Lawrence and He,  Bian and Horvath,  Sean and Ionita,  Monica and Kay,  Jennifer E. and Kim,  Eliot and Kimura,  Noriaki and Kondrashov,  Dmitri and Labe,  Zachary M. and Lee,  WooSung and Lee,  Younjoo J. and Li,  Cuihua and Li,  Xuewei and Lin,  Yongcheng and Liu,  Yanyun and Maslowski,  Wieslaw and Massonnet,  Fran\c{c}ois and Meier,  Walter N. and Merryfield,  William J. and Myint,  Hannah and Navarro,  Juan C. Acosta and Petty,  Alek and Qiao,  Fangli and Schr\"{o}der,  David and Schweiger,  Axel and Shu,  Qi and Sigmond,  Michael and Steele,  Michael and Stroeve,  Julienne and Sun,  Nico and Tietsche,  Steffen and Tsamados,  Michel and Wang,  Keguang and Wang,  Jianwu and Wang,  Wanqiu and Wang,  Yiguo and Wang,  Yun and Williams,  James and Yang,  Qinghua and Yuan,  Xiaojun and Zhang,  Jinlun and Zhang,  Yongfei},
  year = {2024},
  month = jul,
  pages = {E1170–E1203}
}

@online{AIFSThalassa,
  author       = {    Jakob Schloer and
    Steffen Tietsche and
    Christopher Roberts and
    Simon Lang and
    Lorenzo Zampieri and
    Sara Hahner and
    Rachel Furner and
    Mariana Clare and
    Gareth Jones},
  title        = {{AIFS team page for AI Weather Quest, Model summary for AIFS-Thalassa}},
  organization = {European Centre for Medium-Range Weather Forecasts (ECMWF)},
  year         = {2025},
  url          = {https://aiweatherquest.ecmwf.int/team/aifs/},
  note         = {Accessed: 2026-01-29},
}

@article{Loegel2025,
doi = {10.1088/3049-4753/adf649},
url = {https://doi.org/10.1088/3049-4753/adf649},
year = {2025},
month = {aug},
publisher = {IOP Publishing},
volume = {1},
number = {1},
pages = {010701},
author = {Loegel, Olga and Talib, Joshua and Vitart, Frederic and Hoffmann, Jörn and Chantry, Matthew},
title = {The AI Weather Quest: an international competition for sub-seasonal forecasting with AI},
journal = {Machine Learning: Earth},

}

@misc{Huang2025,
  doi = {10.48550/ARXIV.2506.03210},
  url = {https://arxiv.org/abs/2506.03210},
  author = {Huang,  Qiusheng and Niu,  Yuan and Zhong,  Xiaohui and Guo,  Anboyu and Chen,  Lei and Zhang,  Dianjun and Zhang,  Xuefeng and Li,  Hao},
  title = {FuXi-Ocean: A Global Ocean Forecasting System with Sub-Daily Resolution},
  publisher = {arXiv},
  year = {2025},
  copyright = {arXiv.org perpetual,  non-exclusive license}
}

@article{Cui2025,
  title = {Forecasting the eddying ocean with a deep neural network},
  volume = {16},
  ISSN = {2041-1723},
  url = {http://dx.doi.org/10.1038/s41467-025-57389-2},
  DOI = {10.1038/s41467-025-57389-2},
  number = {1},
  journal = {Nature Communications},
  publisher = {Springer Science and Business Media LLC},
  author = {Cui,  Yingzhe and Wu,  Ruohan and Zhang,  Xiang and Zhu,  Ziqi and Liu,  Bo and Shi,  Jun and Chen,  Junshi and Liu,  Hailong and Zhou,  Shenghui and Su,  Liang and Jing,  Zhao and An,  Hong and Wu,  Lixin},
  year = {2025},
  month = mar 
}

@inproceedings{
aouni2025oceanbench,
title={{OceanBench: A Benchmark for Data-Driven Global Ocean Forecasting systems}},
author={El Aouni, Anass and Gaudel, Quentin and Johnson, Juan Emmanuel and Regnier, Charly and Le Sommer, Julien and Van Gennip and Fablet, Ronan and Drevillon, Marie and Drillet, Yann and Le Traon, Pierre Yves},
booktitle={The Thirty-ninth Annual Conference on Neural Information Processing Systems Datasets and Benchmarks Track},
year={2025},
url={https://openreview.net/forum?id=wZGe1Kqs8G}
}

@article{Dheeshjith2025,
  title = {{Samudra: An AI Global Ocean Emulator for Climate}},
  volume = {52},
  ISSN = {1944-8007},
  url = {http://dx.doi.org/10.1029/2024GL114318},
  DOI = {10.1029/2024gl114318},
  number = {10},
  journal = {Geophysical Research Letters},
  publisher = {American Geophysical Union (AGU)},
  author = {Dheeshjith,  Surya and Subel,  Adam and Adcroft,  Alistair and Busecke,  Julius and Fernandez‐Granda,  Carlos and Gupta,  Shubham and Zanna,  Laure},
  year = {2025},
  month = may 
}

@article{Guo2025,
  title = {Data-driven global ocean modeling for seasonal to decadal prediction},
  volume = {11},
  ISSN = {2375-2548},
  url = {http://dx.doi.org/10.1126/sciadv.adu2488},
  DOI = {10.1126/sciadv.adu2488},
  number = {33},
  journal = {Science Advances},
  publisher = {American Association for the Advancement of Science (AAAS)},
  author = {Guo,  Zijie and Lyu,  Pumeng and Ling,  Fenghua and Bai,  Lei and Luo,  Jing-Jia and Boers,  Niklas and Yamagata,  Toshio and Izumo,  Takeshi and Cravatte,  Sophie and Capotondi,  Antonietta and Ouyang,  Wanli},
  year = {2025},
  month = aug 
}

@misc{Gregory2026,
  doi = {10.48550/ARXIV.2603.12449},
  url = {https://arxiv.org/abs/2603.12449},
  author = {Gregory,  William and Bushuk,  Mitchell and Duncan,  James and Wu,  Elynn and Subel,  Adam and Clark,  Spencer K. and Hurlin,  Bill and Watt-Meyer,  Oliver and Adcroft,  Alistair and Bretherton,  Chris and Zanna,  Laure},
  title = {{FloeNet: A mass-conserving global sea ice emulator that generalizes across climates}},
  publisher = {arXiv},
  year = {2026},
  copyright = {Creative Commons Attribution 4.0 International}
}

@article{Finn2024,
  title = {{Generative Diffusion for Regional Surrogate Models From Sea‐Ice Simulations}},
  volume = {16},
  ISSN = {1942-2466},
  url = {http://dx.doi.org/10.1029/2024MS004395},
  DOI = {10.1029/2024ms004395},
  number = {10},
  journal = {Journal of Advances in Modeling Earth Systems},
  publisher = {American Geophysical Union (AGU)},
  author = {Finn,  Tobias Sebastian and Durand,  Charlotte and Farchi,  Alban and Bocquet,  Marc and Rampal,  Pierre and Carrassi,  Alberto},
  year = {2024},
  month = oct 
}

@article{Hakim2024,
  title={Dynamical tests of a deep learning weather prediction model},
  author={Hakim, Gregory J and Masanam, Sanjit},
  journal={Artificial Intelligence for the Earth Systems},
  volume={3},
  number={3},
  year={2024},
  publisher={American Meteorological Society}
}

@misc{Clark2024,
  doi = {10.48550/ARXIV.2412.04418},
  url = {https://arxiv.org/abs/2412.04418},
  author = {Clark,  Spencer K. and Watt-Meyer,  Oliver and Kwa,  Anna and McGibbon,  Jeremy and Henn,  Brian and Perkins,  W. Andre and Wu,  Elynn and Harris,  Lucas M. and Bretherton,  Christopher S.},
  title = {ACE2-SOM: Coupling an ML atmospheric emulator to a slab ocean and learning the sensitivity of climate to changed CO$_2$},
  publisher = {arXiv},
  year = {2024},
  copyright = {Creative Commons Attribution 4.0 International}
}

@misc{Duncan2025,
  doi = {10.48550/ARXIV.2509.12490},
  url = {https://arxiv.org/abs/2509.12490},
  author = {Duncan,  James P. C. and Wu,  Elynn and Dheeshjith,  Surya and Subel,  Adam and Arcomano,  Troy and Clark,  Spencer K. and Henn,  Brian and Kwa,  Anna and McGibbon,  Jeremy and Perkins,  W. Andre and Gregory,  William and Fernandez-Granda,  Carlos and Busecke,  Julius and Watt-Meyer,  Oliver and Hurlin,  William J. and Adcroft,  Alistair and Zanna,  Laure and Bretherton,  Christopher},
  title = {{SamudrACE: Fast and Accurate Coupled Climate Modeling with 3D Ocean and Atmosphere Emulators}},
  publisher = {arXiv},
  year = {2025},
  copyright = {Creative Commons Attribution 4.0 International}
}

@article{CresswellClay2025,
  title = {{A Deep Learning Earth System Model for Efficient Simulation of the Observed Climate}},
  volume = {6},
  ISSN = {2576-604X},
  url = {http://dx.doi.org/10.1029/2025AV001706},
  DOI = {10.1029/2025av001706},
  number = {4},
  journal = {AGU Advances},
  publisher = {American Geophysical Union (AGU)},
  author = {Cresswell‐Clay,  Nathaniel and Liu,  Bowen and Durran,  Dale R. and Liu,  Zihui and Espinosa,  Zachary I. and Moreno,  Raul A. and Karlbauer,  Matthias},
  year = {2025},
  month = aug 
}

@article{Browne2025,
  title = {Sea ice data assimilation in {ORAS6}},
  url = {http://dx.doi.org/10.5194/egusphere-2025-3991},
  DOI = {10.5194/egusphere-2025-3991},
  publisher = {Copernicus GmbH},
  author = {Browne,  Philip and de Boisseson,  Eric and Keeley,  Sarah and Pelletier,  Charles and Zuo,  Hao},
  year = {2025},
  month = sep 
}

@article{Hell2025,
  title = {{A Particle‐in‐Cell Wave Model for Efficient Sea‐State Estimates in Earth System Models—PiCLES}},
  volume = {17},
  ISSN = {1942-2466},
  url = {http://dx.doi.org/10.1029/2025MS005221},
  DOI = {10.1029/2025ms005221},
  number = {8},
  journal = {Journal of Advances in Modeling Earth Systems},
  publisher = {American Geophysical Union (AGU)},
  author = {Hell,  Momme and Fox‐Kemper,  Baylor and Chapron,  Bertrand},
  year = {2025},
  month = aug 
}

@article{Beraki2015,
  title = {On the comparison between seasonal predictive skill of global circulation models: Coupled versus uncoupled},
  volume = {120},
  ISSN = {2169-8996},
  url = {http://dx.doi.org/10.1002/2015JD023839},
  DOI = {10.1002/2015jd023839},
  number = {21},
  journal = {Journal of Geophysical Research: Atmospheres},
  publisher = {American Geophysical Union (AGU)},
  author = {Beraki,  Asmerom F. and Landman,  Willem A. and DeWitt,  David},
  year = {2015},
  month = nov 
}

@article{Graham2005,
  title = {{A performance comparison of coupled and uncoupled versions of the Met Office seasonal prediction general circulation model}},
  volume = {57},
  ISSN = {1600-0870},
  url = {http://dx.doi.org/10.3402/tellusa.v57i3.14666},
  DOI = {10.3402/tellusa.v57i3.14666},
  number = {3},
  journal = {Tellus A: Dynamic Meteorology and Oceanography},
  publisher = {Stockholm University Press},
  author = {Graham,  Richard J. and Gordon,  M. and McLean,  P. J. and Ineson,  S. and Huddleston,  M. R. and Davey,  M. K. and Brookshaw,  A. and Barnes,  R. T. H.},
  year = {2005},
  month = Jan,
  pages = {320}
}

@article{Chrust2024,
  title = {Impact of ensemble‐based hybrid background‐error covariances in ECMWF’s next‐generation ocean reanalysis system},
  volume = {151},
  ISSN = {1477-870X},
  url = {http://dx.doi.org/10.1002/qj.4914},
  DOI = {10.1002/qj.4914},
  number = {767},
  journal = {Quarterly Journal of the Royal Meteorological Society},
  publisher = {Wiley},
  author = {Chrust,  Marcin and Weaver,  Anthony T. and Browne,  Philip and Zuo,  Hao and Balmaseda,  Magdalena Alonso},
  year = {2024},
  month = dec 
}

@article{Tietsche2011,
  title = {Recovery mechanisms of Arctic summer sea ice: RECOVERY MECHANISMS OF ARCTIC SUMMER SEA ICE},
  volume = {38},
  ISSN = {0094-8276},
  url = {http://dx.doi.org/10.1029/2010GL045698},
  DOI = {10.1029/2010gl045698},
  number = {2},
  journal = {Geophysical Research Letters},
  publisher = {American Geophysical Union (AGU)},
  author = {Tietsche,  Steffen and Notz,  Dirk and Jungclaus,  Johann H. and Marotzke,  Jochem},
  year = {2011},
  month = jan
}

@article{Schrder2007,
  title = {Impact of instantaneous sea ice removal in a coupled general circulation model},
  volume = {34},
  ISSN = {1944-8007},
  url = {http://dx.doi.org/10.1029/2007GL030253},
  DOI = {10.1029/2007gl030253},
  number = {14},
  journal = {Geophysical Research Letters},
  publisher = {American Geophysical Union (AGU)},
  author = {Schr\"{o}der,  David and Connolley,  William M.},
  year = {2007},
  month = jul 
}

@unpublished{Bidlot2026prep,
  author = {Jean-Raymond Bidlot and Josh Kousal and Saleh Abdalla},
  title  = {{Wave Hindcasts for ERA6 Preparation and Training Data Driven Models}},
  note   = {Manuscript in preparation},
  year   = {2026}
}

@article{Price2023, 
title={{G}en{C}ast: Diffusion-based ensemble forecasting for medium-range weather}, 
author={Ilan Price and Alvaro Sanchez-Gonzalez and Ferran Alet and Timo Ewalds and Andrew El-Kadi and Jacklynn Stott and Shakir Mohamed and Peter Battaglia and Remi Lam and Matthew Willson}, 
year={2023}, 
journal={arXiv preprint arXiv:2312.15796} }

@misc{DestinE2025Waves,
  author       = {Sara Hahner and Jean Bidlot and Josh Kousal and Lorenzo Zampieri and Christian Lessig and Matthew Chantry},
  title        = {Towards an ML-based Earth System Model: Waves},
  howpublished = {Destination Earth (DestinE) blog},
  year         = {2025},
  url          = {https://destine.ecmwf.int/news/destine-blog-towards-an-ml-based-earth-system-model-waves/},
  note         = {Accessed: 17 April 2026}
}

@misc{DestinE2026SeaIce,
  author       = {Lorenzo Zampieri and Sara Hahner and Rachel Furner and Sarah Keeley and Matthew Chantry},
  title        = {{Towards an ML-based Earth System Model: Sea Ice}},
  howpublished = {{Destination Earth (DestinE) blog}},
  year         = {2026},
  url          = {https://destine.ecmwf.int/news/destine-blog-towards-an-ml-based-earth-system-model-sea-ice/},
  note         = {Accessed: 17 April 2026}
}

@misc{Bouallegue2023,
  author = {Zied Ben Bouallegue and Mihai Alexe and Matthew Chantry and Mariana Clare and Jesper Dramsch and Simon Lang and Christian Lessig and Linus Magnusson and Ana Prieto Nemesio and Florian Pinault and Baudouin Raoult and Steffen Tietsche},
  title = {A new ML model in the ECMWF web charts},
  year = {2023},
  institution = {European Centre for Medium-Range Weather Forecasts (ECMWF)},
  url = {https://www.ecmwf.int/en/about/media-centre/aifs-blog/2023/new-ml-model-ecmwf-web-charts},
  doi = {10.21957/4f6e48352d}
}

@article{Gross2018,
  title = {{Physics–Dynamics Coupling in Weather,  Climate,  and Earth System Models: Challenges and Recent Progress}},
  volume = {146},
  ISSN = {1520-0493},
  url = {http://dx.doi.org/10.1175/MWR-D-17-0345.1},
  DOI = {10.1175/mwr-d-17-0345.1},
  number = {11},
  journal = {Monthly Weather Review},
  publisher = {American Meteorological Society},
  author = {Gross,  Markus and Wan,  Hui and Rasch,  Philip J. and Caldwell,  Peter M. and Williamson,  David L. and Klocke,  Daniel and Jablonowski,  Christiane and Thatcher,  Diana R. and Wood,  Nigel and Cullen,  Mike and Beare,  Bob and Willett,  Martin and Lemarié,  Florian and Blayo,  Eric and Malardel,  Sylvie and Termonia,  Piet and Gassmann,  Almut and Lauritzen,  Peter H. and Johansen,  Hans and Zarzycki,  Colin M. and Sakaguchi,  Koichi and Leung,  Ruby},
  year = {2018},
  month = Nov,
  pages = {3505–3544}
}

@article{Schller2025,
  title = {{Quantifying coupling errors in atmosphere-ocean-sea ice models: A study of iterative and non-iterative approaches  in the EC-Earth AOSCM}},
  volume = {18},
  ISSN = {1991-9603},
  url = {http://dx.doi.org/10.5194/gmd-18-9167-2025},
  DOI = {10.5194/gmd-18-9167-2025},
  number = {22},
  journal = {Geoscientific Model Development},
  publisher = {Copernicus GmbH},
  author = {Sch\"{u}ller,  Valentina and Lemarié,  Florian and Birken,  Philipp and Blayo,  Eric},
  year = {2025},
  month = Nov,
  pages = {9167–9187}
}

@article{Gurmy2005,
  title = {{Actual and potential skill of seasonal predictions using the CNRM contribution to DEMETER: coupled versus uncoupled model}},
  volume = {57},
  ISSN = {1600-0870},
  url = {http://dx.doi.org/10.3402/tellusa.v57i3.14655},
  DOI = {10.3402/tellusa.v57i3.14655},
  number = {3},
  journal = {Tellus A: Dynamic Meteorology and Oceanography},
  publisher = {Stockholm University Press},
  author = {Guérémy,  Jean-Fran\c{c}ois and Déqué,  Michel and Braun,  Alain and Piedelièvre,  Jean-Philippe},
  year = {2005},
  month = Jan,
  pages = {308}
}

@article{Vellinga2020,
  title = {{Evaluating Benefits of Two-Way Ocean–Atmosphere Coupling for Global NWP Forecasts}},
  volume = {35},
  ISSN = {1520-0434},
  url = {http://dx.doi.org/10.1175/WAF-D-20-0035.1},
  DOI = {10.1175/waf-d-20-0035.1},
  number = {5},
  journal = {Weather and Forecasting},
  publisher = {American Meteorological Society},
  author = {Vellinga,  Michael and Copsey,  Dan and Graham,  Tim and Milton,  Sean and Johns,  Tim},
  year = {2020},
  month = Oct,
  pages = {2127–2144}
}

@article{Brassington2015,
  title = {Progress and challenges in short- to medium-range coupled prediction},
  volume = {8},
  ISSN = {1755-8778},
  url = {http://dx.doi.org/10.1080/1755876X.2015.1049875},
  DOI = {10.1080/1755876x.2015.1049875},
  number = {sup2},
  journal = {Journal of Operational Oceanography},
  publisher = {Informa UK Limited},
  author = {Brassington,  G.B. and Martin,  M.J. and Tolman,  H.L. and Akella,  S. and Balmeseda,  M. and Chambers,  C.R.S. and Chassignet,  E. and Cummings,  J.A. and Drillet,  Y. and Jansen,  P.A.E.M. and Laloyaux,  P. and Lea,  D. and Mehra,  A. and Mirouze,  I. and Ritchie,  H. and Samson,  G. and Sandery,  P.A. and Smith,  G.C. and Suarez,  M. and Todling,  R.},
  year = {2015},
  month = Aug,
  pages = {s239–s258}
}

@Article{Berthou2025,
AUTHOR = {Berthou, S. and Siddorn, J. and Fraser-Leonhardt, V. and Le Traon, P.-Y. and Hoteit, I.},
TITLE = {Towards Earth system modelling: coupled ocean forecasting},
BOOKTITLE = {Ocean prediction: present status and state of the art (OPSR)},
EDITOR = {Enrique Álvarez Fanjul, Stefania Angela Ciliberti, Jay Pearlman, Kirsten Wilmer-Becker, and Swadhin Behera},
PUBLISHER = {Copernicus Publications},
JOURNAL = {State of the Planet},
VOLUME = {5-opsr},
YEAR = {2025},
PAGES = {20},
URL = {https://sp.copernicus.org/articles/5-opsr/20/2025/},
DOI = {10.5194/sp-5-opsr-20-2025}
}

@article{Berthou2016,
  title = {{Influence of submonthly air–sea coupling on heavy precipitation events in the Western Mediterranean basin}},
  volume = {142},
  ISSN = {1477-870X},
  url = {http://dx.doi.org/10.1002/qj.2717},
  DOI = {10.1002/qj.2717},
  number = {S1},
  journal = {Quarterly Journal of the Royal Meteorological Society},
  publisher = {Wiley},
  author = {Berthou,  Ségolène and Mailler,  Sylvain and Drobinski,  Philippe and Arsouze,  Thomas and Bastin,  Sophie and Béranger,  Karine and Flaounas,  Emmanouil and Lebeaupin Brossier,  Cindy and Somot,  Samuel and Stéfanon,  Marc},
  year = {2016},
  month = Mar,
  pages = {453–471}
}

\newpage

\setcounter{figure}{0}
\renewcommand{\thefigure}{S.\arabic{figure}}
\renewcommand{\theHfigure}{S.\arabic{figure}}

\setcounter{table}{0}
\renewcommand{\thetable}{S.\arabic{table}}
\renewcommand{\theHtable}{S.\arabic{table}}

\section{Supplementary Material}

\subsection{Additional Wave Evaluation}
We provide additional significant wave height forecast evaluation and visualisation for the AIFS Waves and AIFS Marine models.
\begin{figure}[h]
    \centering
    \includegraphics[width=0.5\linewidth]{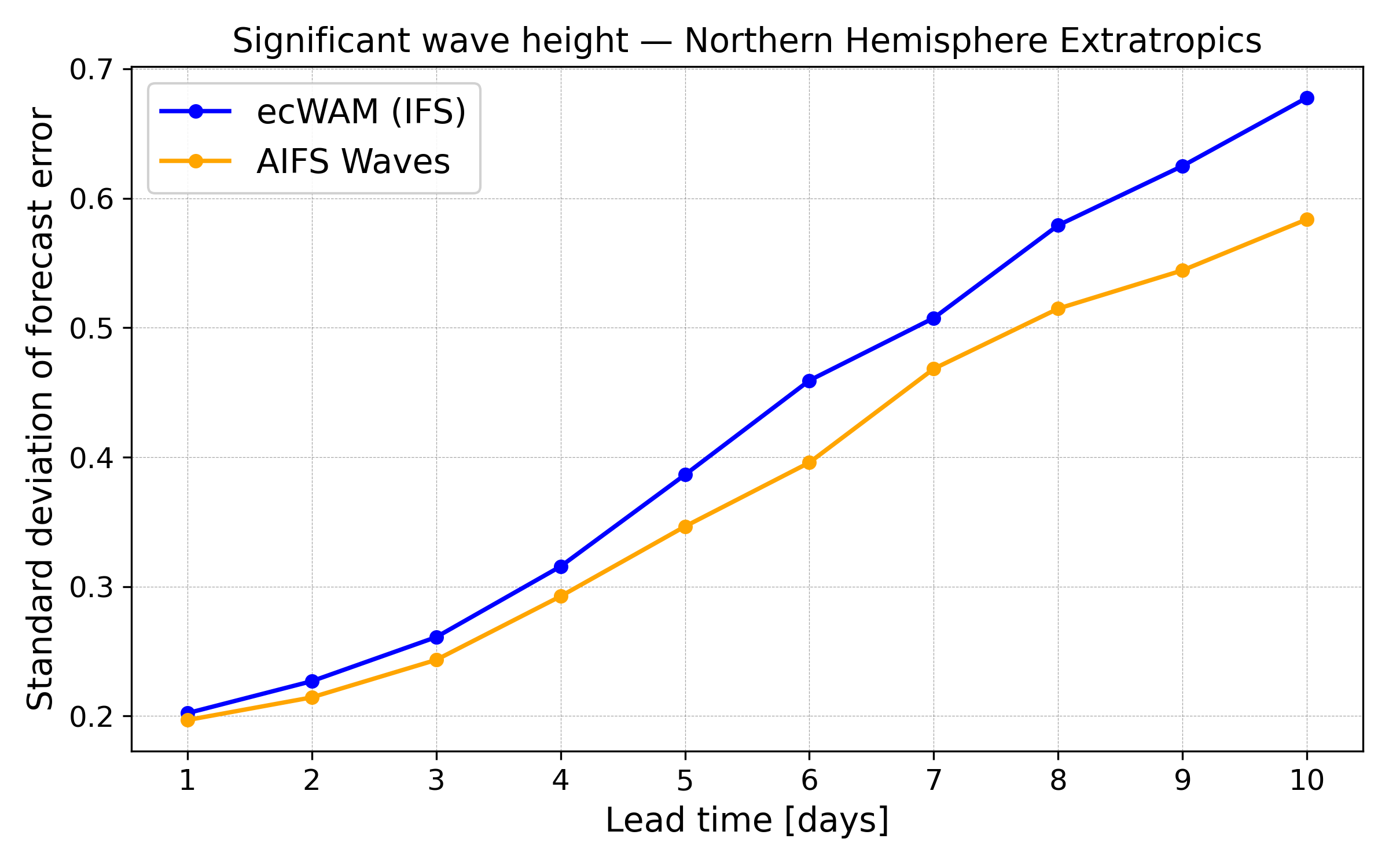}
    \caption{Standard deviation of forecast error for significant wave height forecasts in the northern hemisphere against observations at moored buoys, for May–August 2024 (lower values are better). The joint atmosphere–wave model prototype is in orange, with the physics-based baseline in blue.}
    \label{fig:scores_waves_buoys}
\end{figure}

\begin{figure}[h]
    \centering
    \includegraphics[width=\linewidth]{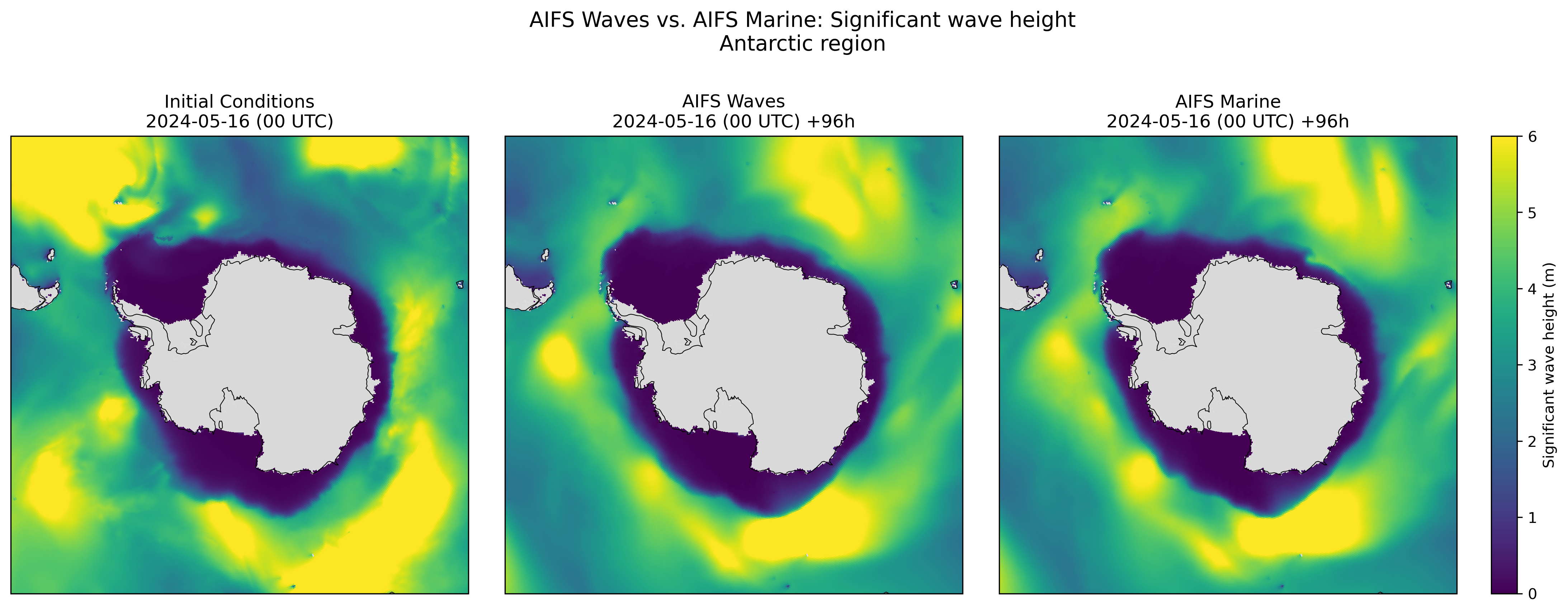}
    \caption{
    Significant wave height (SWH) around Antarctica. Initial conditions on 16 May 2024 (left), 96 h forecast from AIFS Waves (middle), and 96 h forecast from AIFS Marine (right). The sea ice edge appears smoother in AIFS Waves forecasts, whereas it is more sharply defined in AIFS Marine forecasts due to the explicit representation of sea ice.
    }
    \label{fig:sea_ice_edge_waves}
\end{figure}

\begin{figure}[h]
    \centering
    \includegraphics[width=0.55\linewidth]{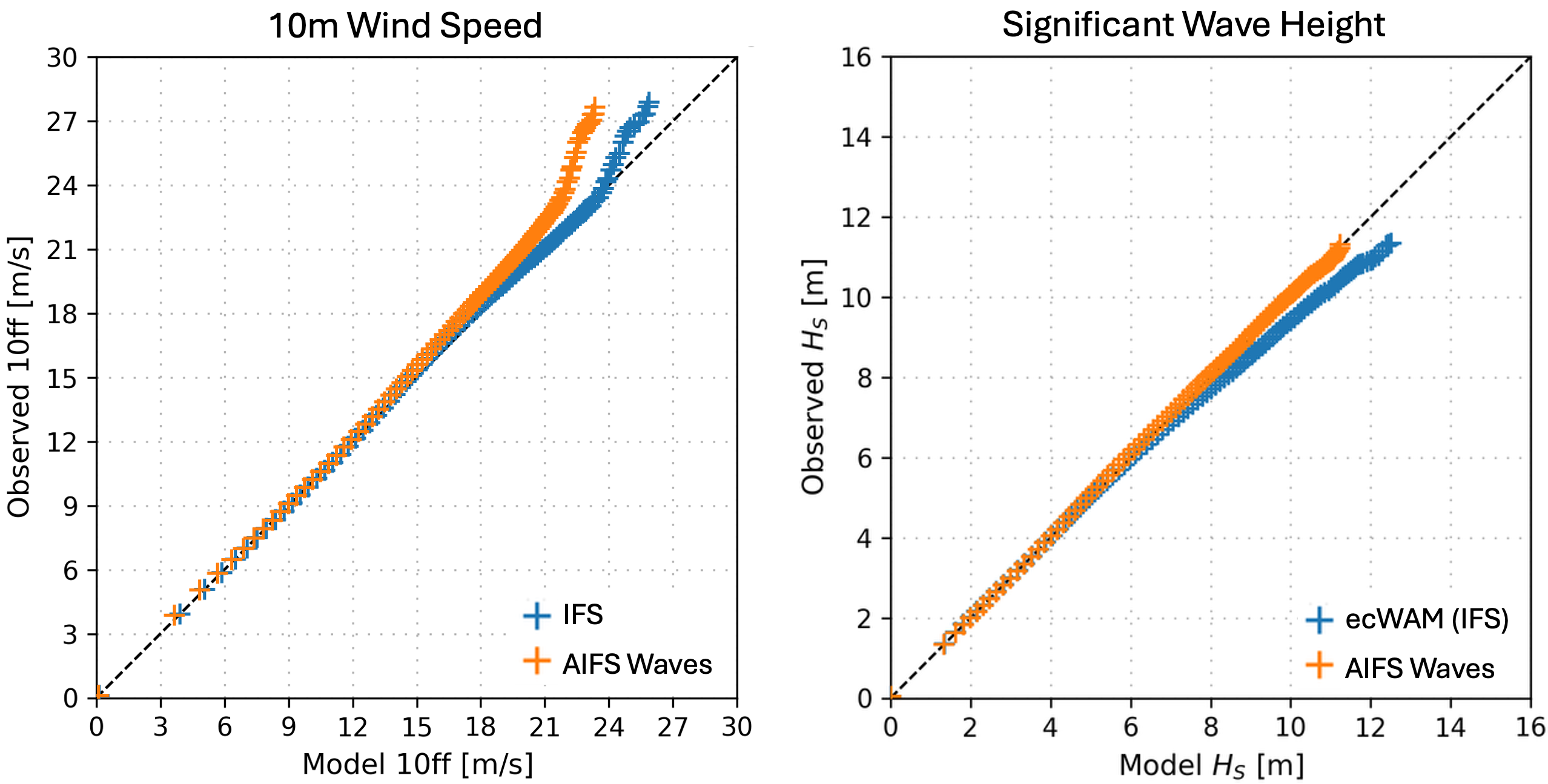}
    \caption{Quantile–quantile diagnostics of 10~m wind speed (10ff, left) and significant wave height (SWH, right) against satellite altimeter observations for May–August 2024 at lead times of 2–3 days. The AIFS Waves model is shown in orange and the corresponding physics-based baseline in blue. Both models are only evaluated for open water conditions, including some lakes. Data-driven wind forecasts exhibit an underrepresentation of strong wind extremes, while the upper tail of the SWH forecast distribution is more accurately captured. }
    \label{fig:wave_extremes}
\end{figure}

\begin{figure}[h]
    \centering
    \includegraphics[width=\linewidth]{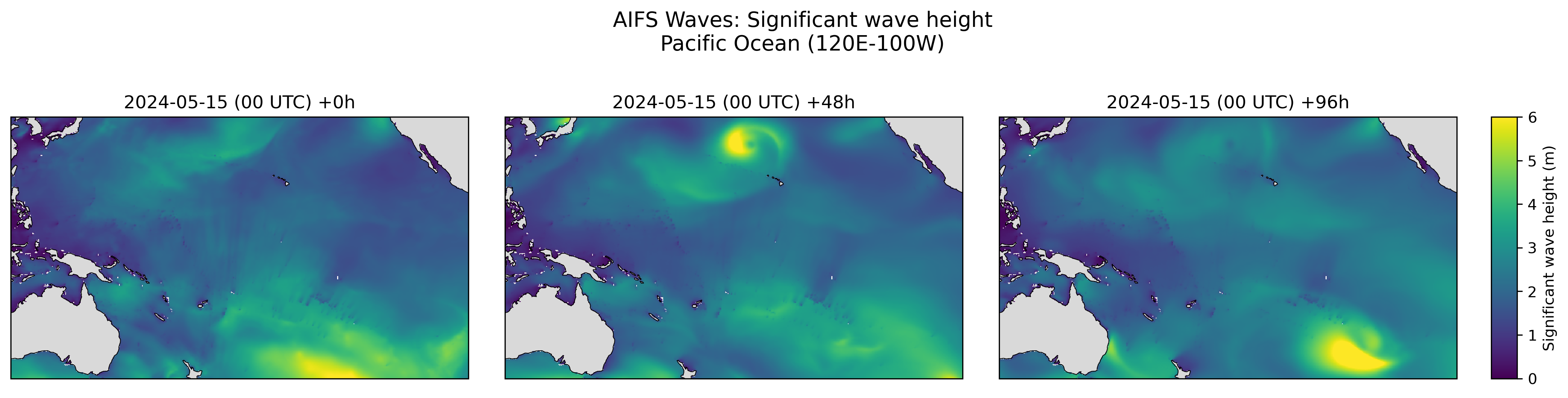}
    \caption{
    Significant wave height (SWH) in the Pacific Ocean showing the interaction of waves with islands. 
    Initial conditions on 15 May 2024 (left), 48 h forecast (middle), and 96 h forecast (right) from the AIFS Waves model. 
    The shadowing effect of islands is clearly visible, although increasingly smoothed at longer lead times.
    }
    \label{fig:waves_islands}
\end{figure}

\FloatBarrier
\subsection{Effect of Using Less Training Years}
\label{sec:effect_training_years}

\begin{figure}[h]
    \centering
    \includegraphics[width=0.43\linewidth]{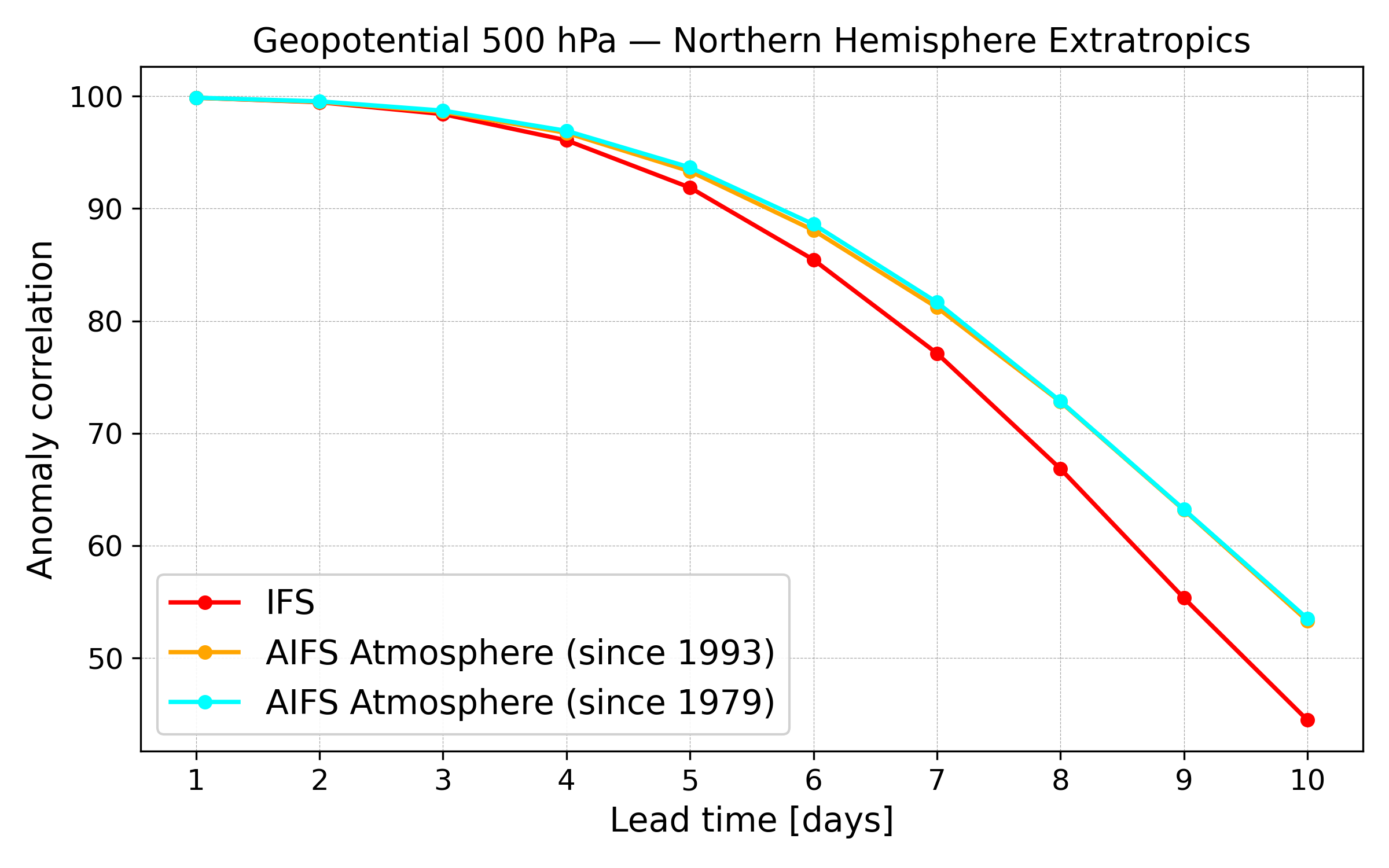}
    \includegraphics[width=0.43\linewidth]{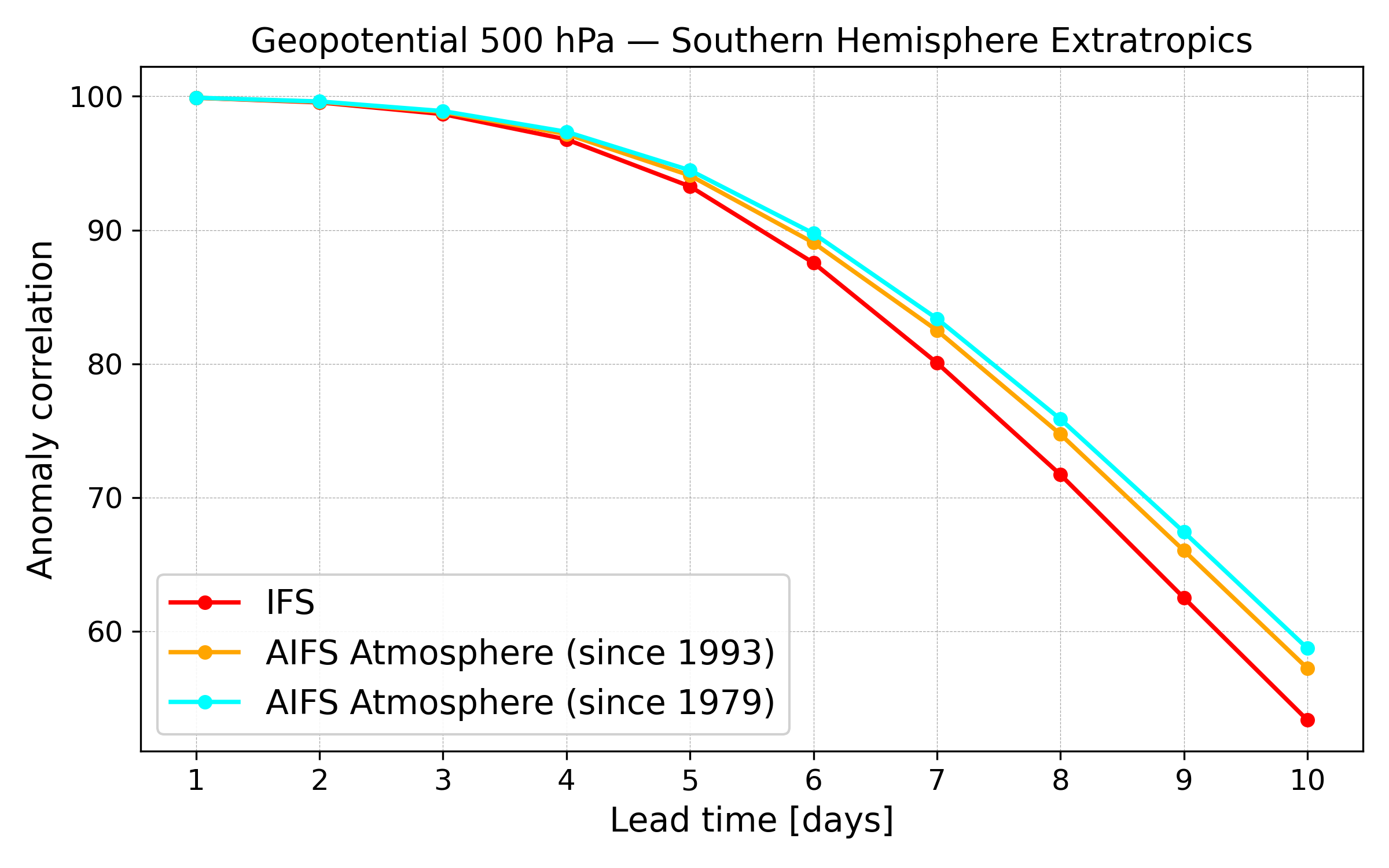}
    \caption{Anomaly correlation skill scores for geopotential at 500hPa in the Northern Hemisphere Extratropics (left) and Southern Hemisphere Extratropics (right). Skill scores computed for 15 June--15 December 2023 against IFS analysis}
    \label{fig:upper_air_scores_years}
\end{figure}

To assess the impact of the reduced pre-training period, we compare models pre-trained on ERA5 data from 1979–2022 and from 1993–2022. 
There are no differences in forecast skill for the northern hemisphere and they are small for the southern hemisphere, see Fig.~\ref{fig:upper_air_scores_years}. 
This is consistent with the training setup, in which the model is subsequently fine-tuned on recent data (2016–2022), which has a strong influence on the final model behaviour.

\FloatBarrier
\subsection{Additional Evaluation of Atmospheric Fields}

\begin{figure}[h]
    \centering
    \includegraphics[width=0.75\linewidth]{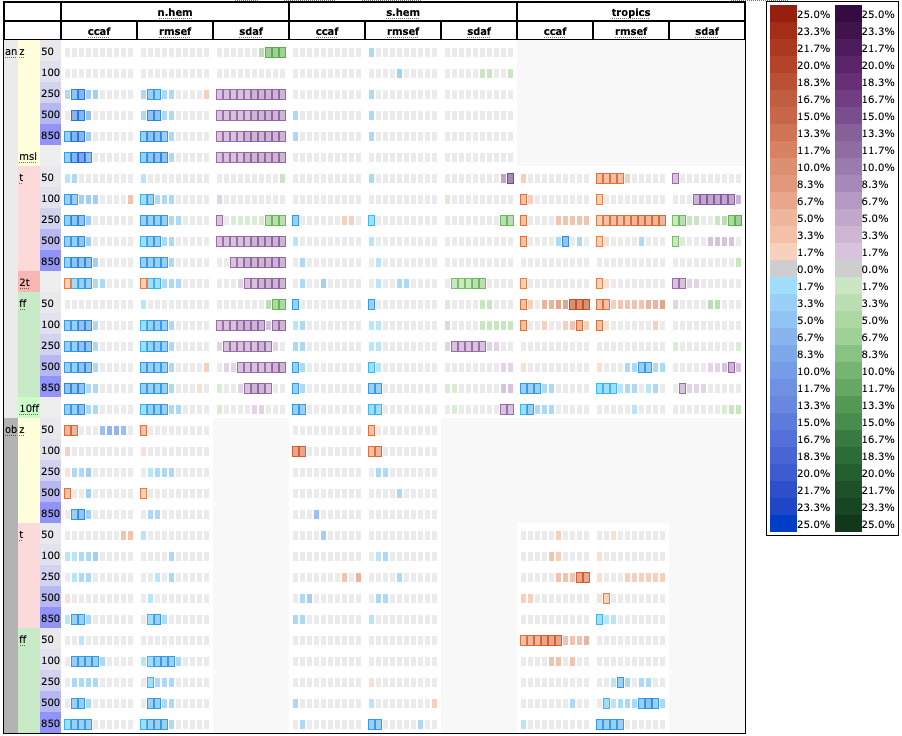}
    \caption{Scorecard comparing forecast scores of AIFS Waves versus AIFS Atmosphere for 15 June--15 December 2023. Forecasts are initialised at 00 and 12 UTC. Relative score changes are shown as function of lead time (day 1 to 10) for northern extra-tropics (n.hem), southern extra-tropics (s.hem) and tropics. Blue colours mark score improvements and red colours score degradations. 
    Purple colours indicate an increase in standard deviation of forecast anomaly, while green colours indicate a reduction. Framed rectangles indicate 95\% significance level. 
    Variables are geopotential (z), temperature (t), wind speed (ff), mean sea level pressure (msl), 2 m temperature (2t), and 10 m wind speed (10ff). Numbers behind variable abbreviations indicate variables on pressure levels (e.g., 500 hPa), and suffix indicates verification against IFS NWP analyses (an) or radiosonde and SYNOP observations (ob). Scores shown are anomaly correlation (ccaf), RMSE (rmsef) and standard deviation of forecast anomaly (sdaf).}
    \label{fig:scorecard_wave}
\end{figure}

\begin{figure}[h]
    \centering
    \includegraphics[width=0.75\linewidth]{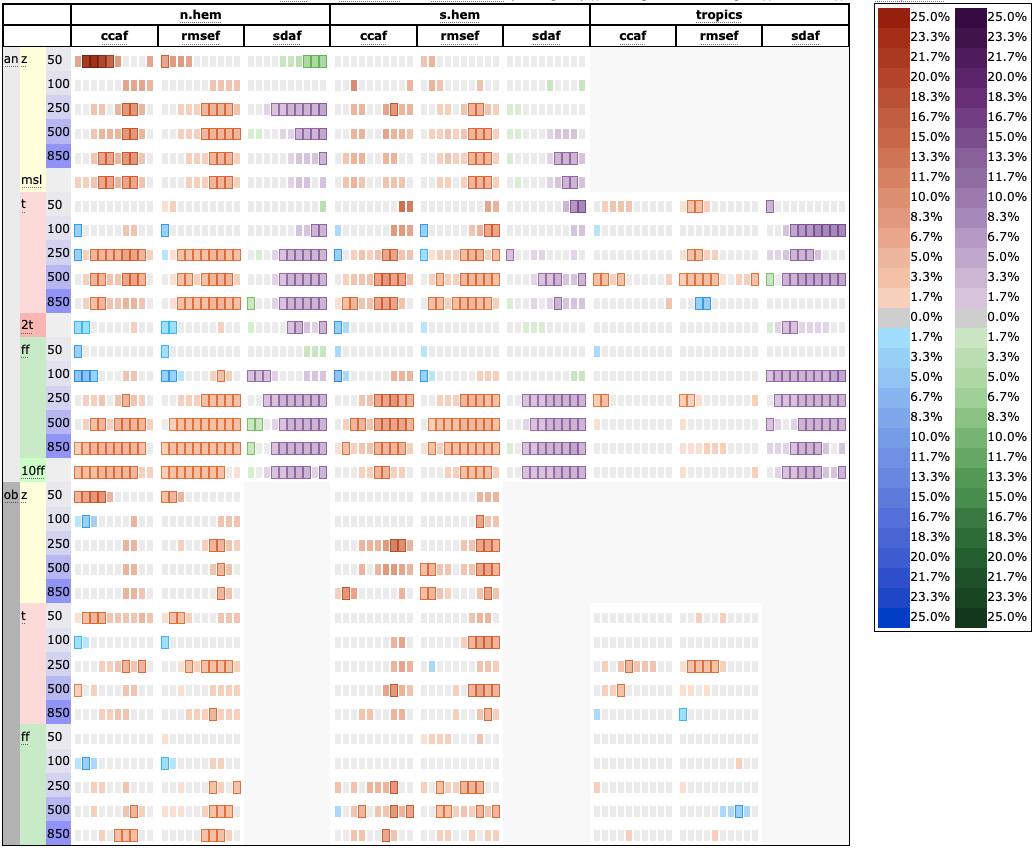}
    \caption{Scorecard comparing forecast scores of AIFS Ocean versus AIFS Atmosphere for 15 June--15 December 2023. For a description of metrics see Fig.~\ref{fig:scorecard_wave}}
    \label{fig:scorecard_ocean}
\end{figure}

\begin{figure}[h]
    \centering
    \includegraphics[width=0.65\linewidth]{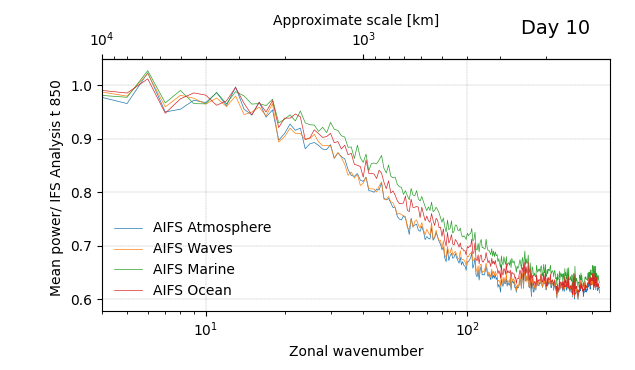}
    \caption{Spectra of 10-day forecasts of temperature at 850 hPa relative to spectra of IFS initial condition. 15th June 2023 until 15th December 2023.}
    \label{fig:spectra}
\end{figure}

We provide additional evaluation results when comparing the different joint model variants to each other.
In Fig.~\ref{fig:scorecard_wave}, we provide a scorecard for the upper air variables of the AIFS Waves against the AIFS, where only the atmosphere is represented, which is neutral in RMSE and forecast activity. 
The scorecard in Fig.~\ref{fig:scorecard_ocean} compares the AIFS Ocean to the AIFS, which shows a degradation for many upper air variables, while at the same time models with an explicit surface ocean representation better capture the spectral distribution, see Fig.~\ref{fig:spectra}.

\FloatBarrier
\clearpage
\subsection{Additional Evaluation on Removing Sea Ice from Initial Conditions}

\begin{figure}[h]
    \centering
    \adjustbox{center}{%
        \includegraphics[width=1.0\linewidth]{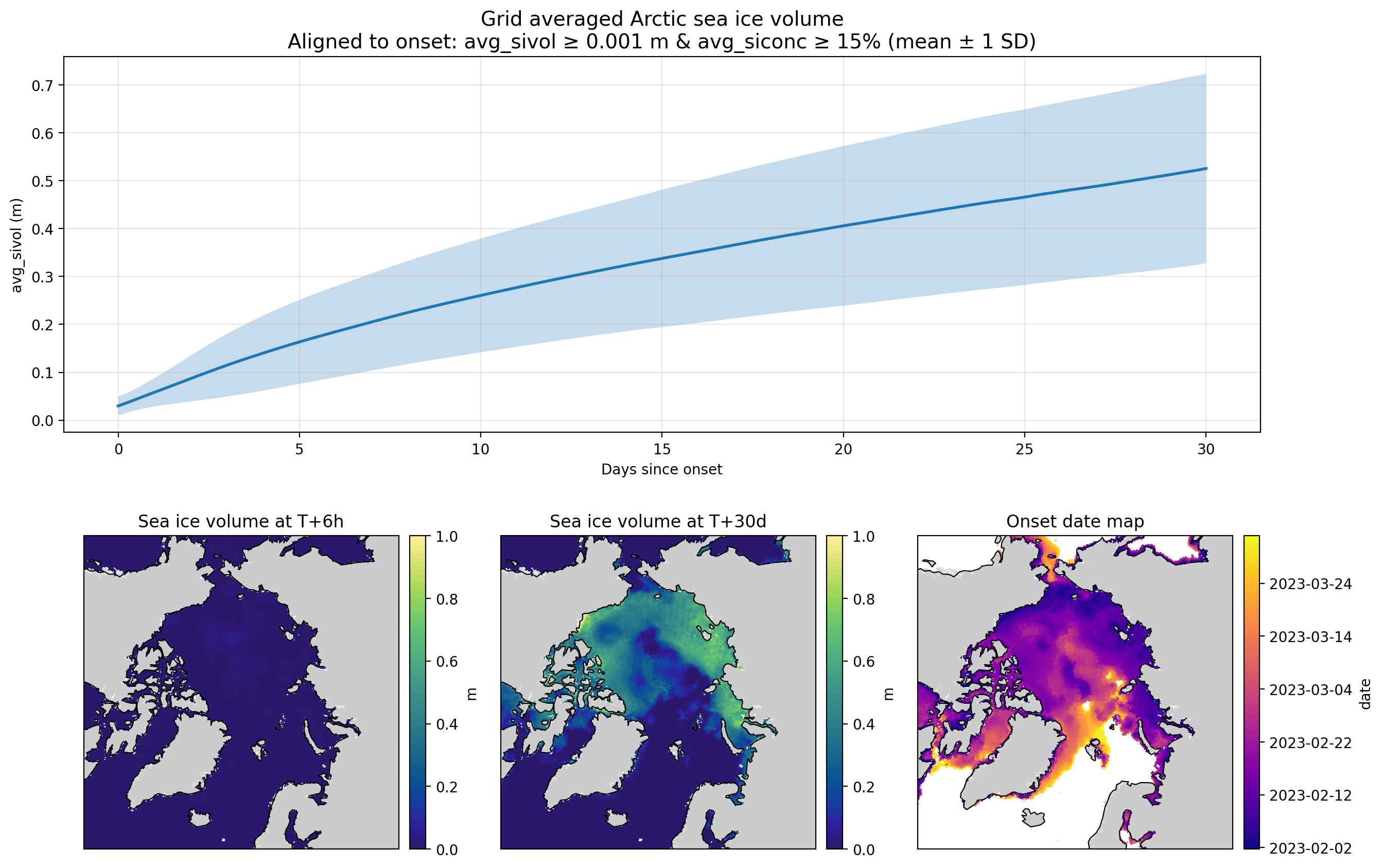}
    }
    \caption{
    Arctic sea ice response in the perturbed forecast initialised on 1 February 2023.
    Top: Grid-point-averaged sea ice volume per unit area, aligned to the local onset of ice formation defined by $\mathrm{avg\_sivol} \ge 0.001\,\mathrm{m}$ and sea ice concentration $\ge 15\%$. 
    The solid line shows the mean across grid points, with shading indicating $\pm 1$ standard deviation, shown for the first 30 days following onset.
    Bottom left: Sea ice volume per unit area at the first forecast step, illustrating the imposed ice-free initial condition.
    Bottom centre: Sea ice volume per unit area after 30 days, showing the spatial pattern of ice recovery.
    Bottom right: Onset date map indicating the first time at which each grid point satisfies the onset criteria.
    }
    \label{fig:seaice_sensitivity_onset}
\end{figure}

\FloatBarrier
\clearpage
\subsection{Variable Selection}

\begin{table}[h]
\centering
\small
\begin{tabular}{p{0.006\linewidth}p{0.41\linewidth}p{0.08\linewidth}p{0.10\linewidth}p{0.1\linewidth}p{0.14\linewidth}}
&\textbf{Variable name} & \textbf{Variable type} & \textbf{Normali-sation} & \textbf{Scaling} & \textbf{Bounding \& \mbox{Postprocessing}}\\
\midrule
\multirow{21}{*}{\rotatebox{90}{Atmosphere}}
& Geopotential & P & Z-score & 12 \\
& Horizontal wind components u and v & P & Z-score & 0.8, 0.5 \\
& Specific humidity & P & Std & 0.6 \\
& Temperature & P & Z-score & 6 \\
& Surface pressure & P & Z-score & 10 \\
& Mean sea-level pressure & P & Z-score & 1 \\
& Skin temperature & P & Z-score & 1 \\
& 2 m temperature & P & Z-score & 1 \\
& 2 m dewpoint temperature & P & Z-score & 0.5 \\
& 10 m horizontal wind components & P & Z-score & 0.5 \\
& Total column water & P & Std & 1 & ReLU(0)\\
& Total precipitation & D & Std & 0.025 & ReLU(0)\\
& Convective precipitation & D & Std (tp) & 0.0025 \\ 
& Land-sea mask & F & None & -- \\
& Orography & F & Max & -- \\
& Standard deviation of sub-grid orography & F & Max & -- \\
& Slope of sub-scale orography & F & Max & -- \\
& Insolation & F & None & -- \\
& Latitude/longitude (cos/sin) & F & None & -- \\
& Time of day / Julian day (cos/sin) & F & None & -- \\
\midrule
\multirow{6}{*}{\rotatebox{90}{Waves}}
& Significant wave height (SWH) & P & Std & 0.5 & ReLU(0)\\
& Mean wave period & P & Std & 0.2 & ReLU(0)\\
& Mean wave direction & P & None & 0.1 & \\
& Coefficient of drag with waves & P & Z-score & 0.01\\
& SWH of waves with periods within 10 and 12~s, 12 and 14~s, 14 and 17~s, 17 and 21~s, 21 and 25~s, and 25 and 30~s & P & Std & 0.1 \mbox{(h2530: 1.0)} & ReLU(0)\\
& Bathymetry & F & None & --\\
\midrule
\multirow{4}{*}{\rotatebox{90}{Ocean}}
& Sea surface temperature & P & Z-score & 50 & ReLU(271.15) \\
& Sea surface height anomaly & P & Z-score & 10 \\
& Sea surface salinity & P & Z-score & 10 & ReLU(0)\\
& Sea surface velocities & P & Z-score & 0.1\\
\midrule
\multirow{5}{*}{\rotatebox{90}{Sea Ice}}
& Sea ice concentration & P & None & 500 & Hardtanh\\
& Sea ice albedo & P & None & 10 & Hardtanh, \newline Zero if siconc = 0\\
& Sea ice volume & P & Std & 10 & ReLU(0), \newline Zero if siconc = 0\\
& Sea ice velocities & P & Std & 0.1 & Zero if siconc = 0\\
& Snow volume over sea ice & P & Std & 10 & ReLU(0), \newline Zero if siconc = 0\\
\end{tabular}
\caption{Variables used in the training of the different AIFS versions, with their short names, level type, variable type, normalisation method, and scaling factors.}
\label{tab:variables}
\end{table}

\FloatBarrier
\subsection{Configurations of the IFS Numerical Model}\label{sec:ifs}

To evaluate the performance of the ML-based AIFS forecasts, we compare against several configurations of the IFS model. The IFS forecasts shown in Figs.~\ref{fig:tc_AIFS_IFS}, \ref{fig:sea_ice_scores}, \ref{fig:ocean_scores}, and \ref{fig:tc_sst_anomaly} are based on a prototype configuration of IFS Cycle 49R2b. This configuration initialises the NEMO4 ocean model from ORAS6 and features tight coupling between the surface ocean and atmosphere, similar to Cycle 50R1, i.e. without the partial coupling present in the operational Cycle 49R1 and earlier versions. Unlike Cycle 50R1, however, this prototype does not include wave attenuation under sea ice. The IFS forecasts shown in Figs.~\ref{fig:upper_air_scores}, \ref{fig:sfc_obs}, and \ref{fig:upper_air_scores_years} are derived from the operational IFS forecast archive. Finally, the numerical wave forecasts shown in Figs.~\ref{fig:scores_waves}, \ref{fig:scores_waves_buoys}, and \ref{fig:wave_extremes} are based on a Cycle 50R1 prototype experiment from the research department, which includes explicit wave attenuation under sea ice. Cycle 50R1 is scheduled for operational implementation at ECMWF on 12 May 2026, replacing Cycle 49R1.

\end{document}